\documentclass[reqno,11pt]{amsart}  

\usepackage{amsmath}
\usepackage{amssymb} 
\usepackage{amsthm}
\usepackage{mathtools}   
\usepackage{graphicx}
\usepackage{graphics}  
\usepackage[pdftex]{color}
\usepackage{color}
\usepackage{verbatim} 
\usepackage[normalem]{ulem} 
\usepackage[mathscr]{eucal}
\usepackage{upgreek}
\usepackage{enumerate} 
\usepackage{color}
\usepackage{dsfont}
\usepackage{tikz}
\usetikzlibrary{decorations.pathreplacing,angles,quotes}
\usepackage{subcaption}
\usepackage{scalerel}

\usepackage[titletoc]{appendix}

\numberwithin{equation}{section}

\newtheorem{theorem}{Theorem}[section] 
\newtheorem{lemma}[theorem]{Lemma} 
\newtheorem{prop}[theorem] {Proposition} 
\newtheorem{cor}[theorem]  {Corollary} 
\newtheorem{remark}[theorem]  {Remark}

\theoremstyle{definition}

\usepackage{float}
\floatstyle{boxed}
\restylefloat{figure}

\DeclareMathAlphabet{\mathpzc}{OT1}{pzc}{m}{it}

\DeclarePairedDelimiter{\abs}{\lvert}{\rvert}

\newcommand{\bL} {\boldsymbol{L}} 
\newcommand{\bM} {\boldsymbol{M}}

\newcommand{\blambda}{\boldsymbol{\lambda}}
\newcommand{\bmu}{\boldsymbol{\mu}}

\newcommand{\bpi}{\boldsymbol{\pi}}

\newcommand{\bK}{\boldsymbol{\mathsf{K}}}

\renewcommand{\L} {\Lambda} 
\renewcommand{\bL}{\boldsymbol{\Lambda}}

\def\d{\delta} 
\newcommand{\e} {\varepsilon}

\newfam\Bbbfam 
\font\tenBbb=msbm10 
\font\sevenBbb=msbm7 
\font\fiveBbb=msbm5 
\textfont\Bbbfam=\tenBbb 
\scriptfont\Bbbfam=\sevenBbb 
\scriptscriptfont\Bbbfam=\fiveBbb

\newcommand{\R}     {\mathbb{R}} 
\newcommand{\Z}     {\mathbb{Z}} 
\newcommand{\N}     {\mathbb{N}} 
\renewcommand{\P}   {\mathbb{P}}

\newcommand{\K} {\mathbb{K}}
\newcommand{\floor}[1]{\left\lfloor #1 \right\rfloor}

\def\1{{\mathchoice {1\mskip-4mu\mathrm l}      
		{1\mskip-4mu\mathrm l} 
		{1\mskip-4.5mu\mathrm l} {1\mskip-5mu\mathrm l}}} 
\newcommand{\ssup}[1] {{\scriptscriptstyle{({#1}})}}

\newtheoremstyle{thm}{2ex}{2ex}{\itshape\rmfamily}{} 
{\bfseries\rmfamily}{}{1.7ex}{} 

\newtheoremstyle{rem}{1.3ex}{1.3ex}{\rmfamily}{} 
{\itshape\rmfamily}{}{1.5ex}{} 

\newenvironment{proofsect}[1] 
{\vskip0.1cm\noindent{\scshape #1.}\hskip0.5cm} 

\def\e{{\rm e}}
\def\dx{{\rm d} x}
\def\dy{{\rm d} y}

\newcommand{\Ccal}   {{\mathcal C }}

\newcommand{\Ncal}   {{\mathcal N }}

\newcommand{\Ucal}   {{\mathcal U }}

\newcommand{\Hscr} {\mathscr{H}}

\newcommand{\Jscr}{\mathscr{J}}
\newcommand{\Kscr}{\mathscr{K}}
\newcommand{\Mscr}{\mathscr{M}}

\newcommand{\m} {{\mathfrak m}}

\newcommand{\ex}{{\rm e}} 
 
\renewcommand{\d}{{\rm d}}

\newcommand{\Li}{\operatorname{Li}\,}
\newcommand{\Leb}{{\rm Leb}}

\newcommand{\Exp}{\mathscr{E}\kern-0.2mm{\operatorname{xp}}}
\newcommand{\Log}{\mathscr{L}\kern-0.2mm{\operatorname{og}}}

\newcommand{\heap}[2]{\genfrac{}{}{0pt}{}{#1}{#2}}

\newcommand\Hyp{{\mathsf H}}

\renewcommand{\Pr} {\mathsf{P}}
\newcommand{\Er} {\mathsf{E}}

\newcommand{\Qr}{\mathsf{Q}}

\newcommand*{\ddprime}{^{\prime\prime}\mkern-1.2mu}

\makeindex

\begin{document}

	\title[\hfill \hfill]
	{An explicit large deviation analysis of the spatial cycle Huang-Yang-Luttinger model}

	
	\author{Stefan Adams and Matthew Dickson}
	
	\thanks{}
	
	\date{}
	
	
	
	
	\maketitle 
	
\begin{abstract}
Here we introduce a family of spatial cycle HYL-type models in which the counter-term only affects cycles longer than some cut-off that diverges in the thermodynamic limit. We derive large deviation principles and explicit pressure expressions for these models, and use the zeroes of the rate functions to study Bose-Einstein condensation.
\end{abstract} 
	
\section{Introduction}
Since London's proposal \cite{London} that the super-fluid phase-transition in ${\rm He}^4 $ is an example of Bose-Einstein condensation (BEC), it has been of interest to know how, in theory, interaction potentials affect the condensation of bosons. London himself conjectured that the momentum-space condensation of bosons is enhanced by spatial repulsion between particles. To address this question, Huang, Yang and Luttinger \cite{HYL57} introduced a model (HYL) of a bosons with a hard-sphere repulsion which displays enhanced condensation. The Hamiltonian for that model is given  in terms of the energy occupation counts (number of particles occupying an eigenstate), and this version  has been studied by the Dublin group in the early 1990s \cite{BLP,BDLPb}. In particular, they show that the density of zero-momentum condensation (ground state energy) in momentum space is a function of the interaction parameters, that is, the mean field parameter and the parameter of the negative counter term in the Hamiltonian. 

\begin{figure} 
		\centering
		\begin{subfigure}[b]{\textwidth}
			\centering
			\begin{tikzpicture}[scale = 2]
			\draw[->] (0,0) -- (4,0) node[below]{$\beta$};
			\draw[->] (0,0) -- (0,3) node[left]{$\mu$};
			\path [fill=gray, draw=gray, line width=0mm] (0.05,3) to [out=270,in=165] (2,0.555) to [out=150,in=280] (0.2,3);
			\draw[very thick] (0.2,3) to [out=280,in=150] (2,0.555) to [out=330,in=180] (4,0.1);
			\draw[very thick] (0.05,3) to [out=270,in=165] (2,0.555);
			\draw[dashed] (2,0.555) -- (2,0) node[below]{$\beta_{\rm t}$};
			\draw[] (3,2) node{$B$};
			\draw[] (1,0.5) node{$A$};
			\draw[] (1,2) node{$\mu_{\rm c}\left(\beta\right)$};
			\draw[] (0.5,1) node{$\mu_{\rm t}\left(\beta\right)$};
			\end{tikzpicture}
			\caption{$d\geq 5$}
		\end{subfigure}
		\begin{subfigure}[b]{\textwidth}
			\centering
			\begin{tikzpicture}[scale = 2]
			\draw[->] (0,0) -- (4,0) node[below]{$\beta$};
			\draw[->] (0,0) -- (0,3) node[left]{$\mu$};
			\path [fill=gray, draw=gray, line width=0mm] (0.2,3) to [out=280,in=180] (4,0.1) -- (4,0.05) to [out=180,in=270] (0.1,3);
			\draw[very thick] (0.2,3) to [out=280,in=180] (4,0.1);
			\draw[very thick] (0.1,3) to [out=270,in=180] (4,0.05);
			\draw[] (3,2) node{$B$};
			\draw[] (1,0.5) node{$A$};
			\draw[] (1,2) node{$\mu_{\rm c}\left(\beta\right)$};
			\draw[] (0.5,1) node{$\mu_{\rm t}\left(\beta\right)$};
			\end{tikzpicture}
			\caption{$d=3,4$}
		\end{subfigure}\\
		\caption{Plots of the $\beta$-$\mu$ phase space for $d\geq 3$. Condensation occurs in region $B$, and does not in region $A$. The transition chemical potential resides in the shaded area, and the transition is discontinuous here. For $d\geq5$ and $\beta\geq\beta_{\rm t}$, the transition is continuous at the boundary. For $d=1,2$ we have a lower bound for the transition like the one appearing in $d=3,4$, but no non-trivial upper bound. Like noted by \cite{Lew86}, there is a transition at finite $\mu$ for all $d\geq 1$.}\label{fig1}
	\end{figure}
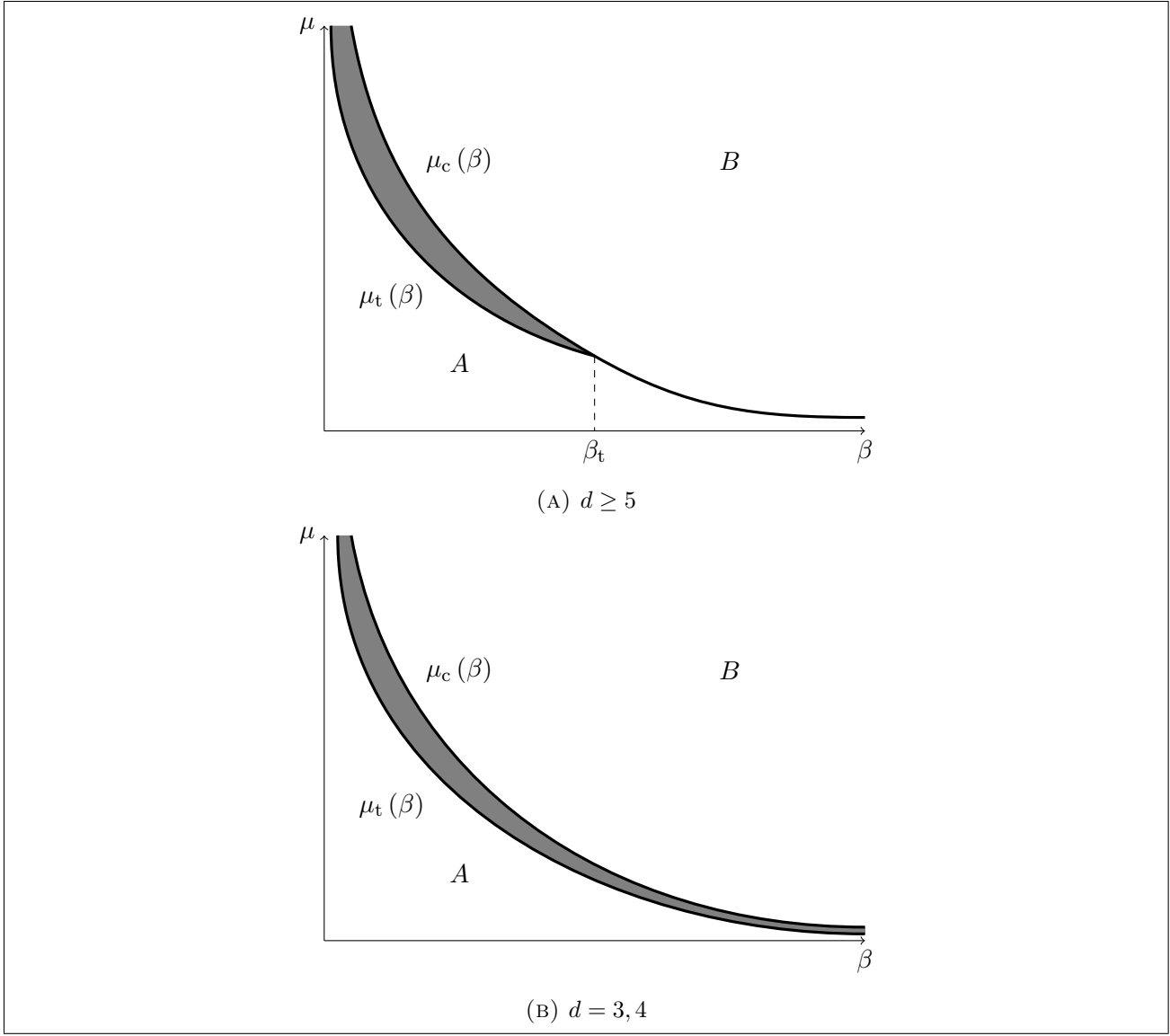

In this paper we study a corresponding model of spatial cycles with HYL-type interaction and we analyse the condensation of parts of the system into `infinitely long cycles'.  The model is given by so-called bosonic cycle counts $\blambda_\L=(\blambda_k)_{k\in\N} $ where $ \abs{\L}\blambda_k $ denotes the number of cycles of length $k$ in finite volume $ \L\Subset\R^d $. Each cycle of length $k$ is a closed continuous path with time horizon $ [0,k\beta] $ and describes $k$ bosons. The bosonic cycle counts are random numbers and their probability weights are obtained from a spatial marked Poisson point process, see \cite{ACK,AD2018}. For the spatial cycle version of the HYL model one needs only the cycle counts and the detailed geometric structures of the cycles do not enter the equation. The Hamiltonian of the spatial cycle HYL model in $ \L\Subset\R^d $ with  interaction parameter $ a>b\ge 0 $ and chemical potential $ \mu\in\R $ is defined as
\begin{equation} \label{Def-H}
\Hyp_\L(\blambda_\L)= \abs{\L}\Big(-(\mu-\alpha)\sum_{k\in\N}k\blambda_k+ \frac{a}{2}\Big(\sum_{k\in\N} k\blambda_k\Big)^2-\frac{b}{2}\sum_{k\ge m_\L} k^2\blambda_k^2\Big),
\end{equation}
where $ (m_\L)_{\L\Subset\R^d} $ is  a sequence with $ m_\L\to\infty $ and $ \frac{m_\L}{\abs{\L}}\to\kappa\in [0,\infty] $ as $ \L\uparrow\R^d $. Here, $ \alpha<0 $ is the chemical potential of the reference process which gives the bosonic cycle count weights for the ideal Bose gas (no interaction), see Section~\ref{sec-setup}.	The Hamiltonian enhances the probability that the particle density, $ \sum_{k\in\N} k\blambda_k $, is concentrated on fewer cycles types. The sequence $ m_\L $ pushes the counter terms to infinitely long cycles and the parameter $ \kappa $ governs how these infinitely long cycles are approached. The most interesting case is when the $ m_\L $ is of order of the thermodynamic scale, the volume $ \L $. Our results below are different for the three possible regimes where either $ \kappa=0, \kappa\in (0,\infty) $, or $ \kappa=\infty $. 
Our main results is a representation of the thermodynamic limit of the pressure via a large deviation principle for the pair empirical particle density $ \bM_\L=(\sum_{k=1}^{m_\L-1} k\blambda_k,\sum_{k\ge m_\L} k\blambda_k) $, see Theorem~\ref{THM1}.  Both the rate function and the pressure depend on the parameter $ \kappa\in[0,\infty] $ characterising the scale on which the counter term cycle lengths grow to infinity. The pressure is given as a variational problem \eqref{pressure(1)}  over  $ \R^2 $, equal to the representation given in \cite{BLP,BDLPb}, where now the second entry refers to the density of particles in `infinitely long cycles'. We obtain  all the zeroes for  the rate functions for all parameter regimes and identify critical parameter with two distinct zeroes, see Theorem~\ref{THM2}. This analysis provides the pressure representation along with the phase transition (BEC) as well as the density of `infinitely long cycles' in the phase transition regime as a function of the chemical potential and the interaction parameter $ a>b$. The relevant parameter is the chemical potential $ \mu $ in the Hamiltonian \eqref{Def-H} and the net chemical potential is $ \mu+\alpha $, and we have some control on the temperature ($\beta $) dependence in Theorem~\ref{thm:mutandmur} leading to phase diagrams in the $ (\mu-\beta) $ - plane, see Figure~ \ref{fig1}. Finally, we compare  the density of `infinetely long cycle' given by the zeroes of the rate functions in Corollary~\ref{Cor-3} with density order parameter defined as the double limit, thermodynamic limit for the density in cycles of length above a  lower  followed by taking the lower bound to infinity, see Theorem~\ref{THM3}. We find that the densities equal as long as $ \kappa\in[0,\infty) $ whereas for $ \kappa=\infty$ the counter terms diverge too quickly to infinity to lead to any non-vanishing density of `infinitely long cycles', see Remark~\ref{rem-final}.

In previous study of the momentum-space model the relevant occupation number  is the lowest index opposite to our spatial cycle model, e.g., in \cite{BLP} the authors need a technical constraint for the counter term up to an index growing to infinity slower than the volume. This technical gap has been closed in \cite{BDLPb} where the all counter term indices are incorporated. However, in those cases for the momentum-space model, the higher indices are not relevant for the zero-momentum condensation, whereas for our model the precise asymptotic of counter terms towards  'infinitely long cycles' play a crucial role.

We define the probability weights for the bosonic cycle counts in Section~\ref{sec-setup}, and show in Section~\ref{Sec-reference} and Appendix~\ref{app-A} how these weights are obtained from a marked Poisson point process description of the ideal Bose gas with free and Dirichlet boundary conditions. Both section can be skipped by the reader at first reading. In Section~\ref{sec-results} we present all our results along with Figure~\ref{fig1} of the phase diagram, and in Section~\ref{sec-proofs} we present all our proofs.
\subsection{Probability weights}\label{sec-setup}  The ideal Bose gas at thermodynamic equilibrium with inverse temperature $ \beta>0 $ and chemical potential $ \alpha< 0 $ defines the probability weights for the bosonic cycle counts when there is no interaction. The cycle counts themselves are actually random functions of marked point configurations $ \omega\in\Omega $ as we will describe in more detail below in Section~\ref{Sec-reference}, that is, $ \blambda_\L=\blambda_\L(\omega) $. For $ x=(x_k)_{k\in\N} \in\ell_1(\R_+) $ with $ \abs{\L}x_k \in\N_0$, we denote by $ \Qr $ the probability distribution of the reference process, that is, the probability that the bosonic cycle counts are equal to $ x$ is given by
\begin{equation*}
\Qr\left(\blambda_{\L} =(x_k)_{k\in\N}\right) = \e^{\abs{\L}\overline{q}^\ssup{\alpha}} \prod_{k\in\N}\frac{\left(\abs{\Lambda}q_k\e^{\beta\alpha k}\right)^{\abs{\Lambda} x_k}}{\left(\abs{\Lambda}x_k\right)!},
\end{equation*}
where 
\begin{equation}\label{q*def}
	\overline{q}^{\ssup{\alpha}}=\sum_{k\in\N}q_k\e^{\beta\alpha k},\qquad\mbox{ with }\quad q_k=\frac{1}{(4\pi\beta)^{d/2}k^{1+d/2}},\qquad k\in\N.
	\end{equation}
	
As shown in \cite{AD2018}, the thermodynamic pressure of the ideal Bose gas  is given as
\begin{equation}
\label{idealPressure}
p_0(\beta,\alpha)=\frac{1}{\beta(4\pi\beta)^{d/2}}\sum_{k=1}^\infty\frac{\ex^{\beta\alpha k}}{k^{1+d/2}}=\frac{1}{\beta}\overline{q}^{\ssup{\alpha}}.
\end{equation}
The spatial cycle HYL model is given by the  Gibbs distribution $ \Pr_\L $ in $ \L $ defined via its Radon-Nikodym density with respect to the reference process $ \Qr_\L $ in $ \L $,
$$
\frac{\d \Pr_\L}{\d\Qr_\L}=\frac{\ex^{-\beta \Hyp_\L}}{\Er\left[\ex^{-\beta \Hyp_\L}\right]}.
$$ 
A vital ingredient in our result is  the free energy of the ideal Bose gas given as the Legendre-Fenchel  transform of the pressure,
	$$
	f_0(\beta,x)=\sup_{\alpha\in\R}\{\alpha x-p_0(\beta,\alpha)\},
	$$
	and, as the pressure $ p_0(\beta,\alpha) $  is differentiable for $ \alpha <0 $ and diverges for $ \alpha>0 $, we obtain the version of the free energy derived in \cite{A08}.

	\subsection{Reference process}\label{Sec-reference}
We define the underlying marked Poisson point reference process in the following as background information, since our results depend solely on the measure $ \Pr_\L $ defined above. 

The reference process is a marked Poisson point process, see \cite{ACK,AD2018}.  The space of marks is defined as
$$
E=\bigcup_{k\in\N} \Ccal_{k,\L}\,, 
$$
where, for $k\in\N$, we denote by $\Ccal_{k,\L}$  the set of continuous functions $f\colon [0,k\beta] \to \mathbb{R}^d$ satisfying $f(0)=f(k\beta)\in\L$,  equipped with the topology of uniform convergence. 
We call the marks {\it cycles}. By  $\ell\colon E\to\N $  we denote the canonical map defined by $\ell(f)=k$ if $f\in \Ccal_{k, \L}$. We  call $\ell(f)$ the {\it length} of $f\in E$. 	
We consider spatial configurations that consist of a locally finite set $\xi\subset\R^d$ of particles, and to each particle $x\in\xi$ we attach a mark $f_x\in E$ satisfying $f_x(0)=x$. Hence a configuration is described by the counting measure $ \omega=\sum_{x\in\xi}\delta_{(x,f_x)} $  on $\R^d\times E$.	
		
\noindent Consider on $\Ccal_{k,\L}$ the canonical Brownian bridge measure
$$
\bmu^{\ssup{k\beta}}_{x,y}(A)=\frac{\P_x(B\in A;B_{k\beta}\in\d y)}{\d y},\qquad A\subset\Ccal_{k,\L}\mbox{ measurable}.
$$
Here $B=(B_t)_{t\in[0,k\beta]}$ is a Brownian motion in $\R^d$ with generator $\Delta$, starting from $x $ under $\P_x$. Then $ \mu^{\ssup{ k\beta}}_{x,y}$ is a regular Borel measure on $\Ccal_{k,\L}$ with total mass equal to the Gaussian density,
$$
\bmu_{x,y}^{\ssup{k \beta}}(\Ccal_{k,\L})=g_{k\beta}(x,y)=\frac {\P_x(B_{k\beta}\in\d y)}{\d y}=(4\pi k\beta)^{-d/2}{\rm e}^{-\frac 1{4 k \beta}|x-y|^2}.
$$ 
Let $ \omega_{\rm P} = \sum_{x \in \xi_{ \rm P}} \delta_{(x,B_x)} $
be a Poisson point process on $\R^d\times E$ with intensity measure equal to $ \nu $, whose projection onto $ \R^d\times\Ccal_{k,\L} $ is equal to 
$$
\nu_k(\d x,\d f)=\frac{1}{k}\Leb(\d x)\otimes\ex^{\beta\alpha k}\mu_{x,x}^{\ssup{k\beta}}(\d f),\qquad k\in\N,\quad \alpha <0.
$$
Alternatively, we can conceive $\omega_{\rm P}$ as a marked Poisson point process  on $\R^d$, based on some Poisson point process $\xi_{\rm P}$ on $\R^d$, and a family $(B_x)_{x\in \xi_{\rm P}}$ of  i.i.d.~marks, given $\xi_{\rm P}$. The intensity of $\xi_{\rm P}$ is  $ \overline{q}^{\ssup{\alpha}} $ \eqref{q*def}.
Denote $ \Qr$ denote the distribution of $\omega_{\rm P}$ and denote by $\Er$ the corresponding expectation. Note that our reference process is a countable superposition of Poisson point processes and, as long as $ \overline{q}^{\ssup{\alpha}} <\infty $ is finite, this reference process is a Poisson point process as well.  The bosonic cycle counts are averages of the random number of points in $ \L $  with mark length equal to $k$, i.e., 
$$
\mathcal{N}_{k}(\omega)=\#\left\{x\in\xi\cap\Lambda \colon  \ell(f_x)=k\right\}, \qquad k\in\N,
$$
and $ \blambda_k(\omega)=\Ncal_k(\omega)/\abs{\L} $.
		
\subsection{Notations}
Throughout the whole text when we write $ \L\uparrow\R^d $ we mean the limit for a sequence of centred finite-volume boxes with unbounded volume. Furthermore,
we write $ f(\lambda)\sim g(\lambda) $ whenever $ \lim_{\lambda\to \infty} \frac{f(\lambda)}{g(\lambda)}=1 $.		
		
\section{Results}\label{sec-results}
	
Our main results concern a complete large deviation analysis and pressure representation of the spatial cycle  HYL-model. Recall that the partition function $ Z_\L(\beta, \alpha) $ for any Hamiltonian, $ \Hyp $,   of marked point configurations can be written as an expectation with respect to the reference process. This follows using the the Feynman-Kac formula for traces with symmetrised initial and terminal conditions (see \cite{BR97,AK08}).
	
\begin{prop}[\text{\cite{ACK,AD2018}}]\label{lem-rewrite}
		Let $\beta\in(0,\infty)$, $ \alpha<0 $ and $ \L\Subset\R^d $. Assume that   $\Hyp_\L\colon\Omega\to \R $  is measurable and bounded from below. Then, 
		\begin{equation*}
		\begin{aligned}
		Z_\L(\beta,\alpha)&=\e^{\abs{\L}\overline{q}^{\ssup{\alpha}}}\Er\left[{\rm e}^{-\beta\Hyp_\L(\omega)}\right].
		\end{aligned}
		\end{equation*}
	\end{prop}
The Hamiltonian of the spatial cycle HYL model is given in \eqref{Def-H} as a function of the bosonic cycle counts $ \blambda(\omega) \in\ell_1(\R_+) $.

The thermodynamic pressure for the spatial cycle HYL-model is 
\begin{equation}\label{pressure-limit}
\begin{aligned}
	p^{\ssup{\kappa}}(\beta,\alpha,\mu)&=\lim_{\L\uparrow\R^d}\frac{1}{\beta\abs{\L}} Z_\L(\beta,\alpha)=p_0(\beta,\alpha)+\lim_{\L\uparrow\R^d}\frac{1}{\beta\abs{\L}}\log \Er\Big[\ex^{-\beta \Hyp_\L}\Big],
	\end{aligned}
	\end{equation}
	where the expectation is with respect to the reference Poisson process and the ideal Bose gas pressure, $p_0\left(\beta,\alpha\right)$, is given by \eqref{idealPressure}, see \cite{AD2018}. The limit on the right hand side of \eqref{pressure-limit} is obtained via a large deviation analysis for the pair empirical particle density $ \bM_\L\colon\Omega\to\R^2 $, 
	\begin{equation*}
	\bM_\L(\omega):=\left(M_\L^{\ssup{1}},M_\L^{\ssup{2}}\right) \quad \mbox{ with } M_\L^{\ssup{1}}=\sum_{k\leq m_\L -1} k\blambda_k \mbox{ and } M_\L^{\ssup{2}}=\sum_{k \ge  m_\L} k\blambda_k,
	\end{equation*}
	distributed under the measure $ \mu_\L^{\ssup{2}}:=\Pr_\L\circ \bM_\L^{-1} $ with $ \Pr_\L $ being the distribution of the Gibbsian point process of the spatial cycle HYL-model in $ \L $, i.e. $\frac{\d \Pr_\L}{\d\Qr_\L}=\ex^{-\beta \Hyp_\L}/\Er\left[\ex^{-\beta \Hyp_\L}\right]$.

	\begin{theorem}[\textbf{Large deviation principles and pressure representation}]\label{THM1}
		Let $ \beta>0$, $\alpha<0$, $\mu\in \R $, and $ a>b>0 $. Let $ (m_\L)_{\L\Subset\R^d} $ be a sequence with $ m_\L\to\infty $ and $ \frac{m_\L}{\abs{\L}}\to\kappa\in [0,\infty] $ as $ \L\uparrow\R^d $, and define $ \K(\kappa):=\R_+\times\left(\{0\}\cup[\kappa,\infty)\right)\subset\R^2_+ $. 
		
		Then the sequence of measures $ (\mu_\L^{\ssup{2}})_{\L\Subset\R^d} $ satisfies the large deviation principle on $ \R^2_+ $ with rate $ \beta\abs{\L} $ and good rate function
		\begin{equation*}
		I^{\ssup{\kappa}}(x,y)=\begin{cases} f_0(\beta,x)-\mu(x+y)+\frac{a}{2}(x+y)^2-\frac{b}{2}y^2+p^{\ssup{\kappa}}(\beta,\alpha,\mu)& \mbox{for } (x,y)\in\K(\kappa),\\
		+\infty &\text{otherwise,}\end{cases}
		\end{equation*}
		where
		\begin{equation}\label{pressure(1)}
		p^{\ssup{\kappa}}(\beta,\alpha,\mu)=\sup_{(x,y)\in\K(\kappa)}\Big\{ \mu(x+y)-\frac{a}{2}\left(x+y\right)^2+\frac{b}{2}y^2 -f_0(\beta,x)\Big\}.
		\end{equation}

	\end{theorem}	
	
	\bigskip
	
	\begin{remark}\label{rem-THM1}

		\begin{enumerate}[(a)]

			\item Since $ \R_+^2=\K(0)\supset \K(\kappa) $, we know $ p^{\ssup{\kappa}}(\beta,\alpha,\mu) \le p^{\ssup{0}} (\beta,\alpha,\mu) $. This is due to the sequence of counter terms forcing the density in the second component of the pair empirical density to be either zero or greater or equal to $ \kappa $.
			\item The case $\kappa=\infty $ (i.e., $ m_\L\gg\abs{\L} $) is different because this condition essentially removes the counter terms in the spatial cycle HYL-model. For finite volumes $ \abs{\L} $, the two smallest permitted values of the random variable $ M_\L^{\ssup{2}} $ are $ 0 $ and $ \frac{m_\L}{\abs{\L}} $. This case then forces all non-zero permitted values of the cycle counts to diverge to infinity, and thus they are naturally unlikely, and the `greater than $ m_\L $' states aren't allowed to support any particle mass. Therefore the spatial cycle HYL-model in this case is essentially equal to the particle mean field model studied for example in \cite{BCMP05,AD2018}.	
			\item 
			The empirical cycle counts $ (\blambda_\L)_{\L\Subset\R^d} $ under the reference process  satisfy, for any $ d\ge 1 $ and $  \alpha<0 $, a large deviation principle with rate $ \beta\abs{\L} $ and rate function
			$$
			I_0\left(x\right) = \begin{cases}
			\sum^\infty_{k=1}x_k\Big(\log\frac{x_k}{q^{\ssup{\alpha}}_k} - 1\Big) + p_0(\beta,\alpha), & \text{for }x\in\ell_1\left(\R_+\right),\\
			+\infty, & \text{otherwise}.
			\end{cases}
			$$ 
			see \cite{AD2018}.   	This large deviation analysis require to deal with the lack of continuity of the particle density  $ D(x)=\sum_{k\in\N} k x_k $ for $ x\in\ell_1(\R) $, see similar problems in \cite{ACK}. The splitting into the pair empirical density allows  to obtain the upper and lower bounds in our large deviation principles  separately with continuous Hamiltonian functions. The large deviation principles in Theorem~\ref{THM1} 	 provide more explicit representations of the pressure allowing a more precise phase transition analysis below in Theorem~\ref{THM2}. 
		\end{enumerate}
		\hfill $ \diamond $
	\end{remark}

	The pressure representations on the right hand side of \eqref{pressure(1)} and their  analysis is the key step in revealing the phase transition phenomenon also known as \textit{Bose-Einstein condensation}.
	
	\begin{cor}
		The right hand side of \eqref{pressure(1)} can, for $ \kappa =0 $, be written as
		\begin{equation*}
		\begin{aligned}		p^{\ssup{0}}(\beta,\alpha,\mu)&=\sup_{x\ge 0} \Big\{\mu x-\frac{a}{2}x^2+\frac{(\mu-ax)_+^2}{2(a-b)}-f_0(\beta,x)\Big\}\\ 
		&=\inf_{\lambda\in\ell_1(\R_+)}\Big\{   I_0(\lambda)-\mu D(\lambda)+\frac{a}{2} D(\lambda)^2-\frac{(\mu-aD(\lambda))_+^2}{2(a-b)}   \Big\}\,,
		\end{aligned}
		\end{equation*}
		where $ I_0 $ is the rate function for the ideal Bose gas and $ D(\lambda)=\sum_{k\in\N} k \lambda_k $ is the density, see  \cite{AD2018}.
	\end{cor}
	
	\begin{proofsect}{Proof}
		The first equality follows from optimising over $ y\ge 0 $, whereas the second can be proved by conditioning on the value of $ D(\lambda) $.
		\qed
	\end{proofsect}

	\medskip

	The analysis of the pressure is related to finding the  zeroes of the rate functions. In the following theorem we identify thus a critical parameter regime for the pressure in terms of the so-called chemical potential $ \mu \in\R $. This analysis is followed by showing that  the critical parameter is  the phase transition point when positive particle density is carried by so-called `infinitely' long cycles, see Theorem~\ref{THM3} below. Recall the critical density of the ideal Bose gas, see \cite{A08,AD2018}, 
	$$
	\varrho_{\rm c}(d)=\begin{cases}\frac{1}{(4\pi\beta)^{d/2}}\zeta\left(\frac{d}{2}\right) & \mbox{ for } d\ge 3,\\
	+\infty & \mbox{ for } d=1,2\,,
	\end{cases}
	$$
	where $ \zeta $ is the Riemann zeta function, $ \zeta\left(\frac{d}{2}\right)=\sum_{k=1}^\infty k^{-\frac{d}{2}} $. For the following results, we define  the function $ s_\beta\colon (0,\infty)\to\R $ by
	\begin{equation*}
	s_\beta(x):=\begin{cases} 0 &\mbox{for } x\ge \varrho_{\rm c}(d),\\ \mbox{unique solution to } p^\prime_0(\beta,s)=x &\mbox{for } 0<x\le \varrho_{\rm c}(d). \end{cases}
	\end{equation*}
	Here and thereafter we let $ p^\prime_0(\beta,s) $ and $ f^\prime_0(\beta,x) $ denote the $ s$ and $x$ partial derivatives respectively. Note that $f_0(\beta,x)=s_\beta(x)x-p_0(\beta,s_\beta(x))$, $f^\prime_0(\beta,x)=s_\beta(x)$, and that $ s_\beta $ is concave everywhere and strictly concave for $ x\le \varrho_{\rm c}(d) $. Define 
	\begin{equation}\label{Def-mu-t}
	\mu_{\rm t}\left(\beta\right) :=\inf_{s<0}\Big\{ap^\prime_0(\beta,s)-\frac{(a-b)}{b} s\Big\} >0,
	\end{equation}
	along with
	\begin{equation*}
	\begin{aligned}
	\tilde{x}_1 & \colon\R\to\R_+, \; \tilde{x}_1(\mu)\mbox{ unique solution to } s_\beta(x)=\mu-ax,\\
	\tilde{x}_2 & \colon[\mu_{\rm t},\infty)\to\R_+, \; \tilde{x}_2(\mu) \mbox{ minimal solution to } s_\beta(x)=-\frac{b}{a-b}(\mu-ax),
	\end{aligned}
	\end{equation*}
	and finally the chemical potential for $ \kappa\in(0,\infty) $, 
	\begin{equation*}
	\mu_{\rm r}\left(\beta,\kappa\right) :=\inf\Big\{ s\ge \mu_{\rm t}\colon \tilde{x}_2(s)\le\frac{1}{a}(s-\kappa(a-b))\Big\} \,\mbox{ and }\, \tilde{x}_3(\mu)=\tilde{x}_1(\mu-a\kappa).
	\end{equation*}
	Denote
	\begin{equation*}
	\Mscr^{\ssup{\kappa}}(\mu):=\left\{\mbox{ zeroes of } I^{\ssup{\kappa}}\right\}\quad\mbox{ for } \kappa\in[0,\infty], \mu\in\R.
	\end{equation*}

	\medskip

	\begin{theorem}[\textbf{Zeroes of the rate functions}]\label{THM2}
		Let $ \beta>0, \alpha<0, \mu\in \R $, $ a>b>0 $, and $ \mu_{\rm c}:= a\varrho_{\rm c}(d) $. Under the same assumptions on the sequence $ (m_\L)_{\L\Subset\Z^d} $ as in  Theorem~\ref{THM1}, the following holds for $ \kappa=0 $, $ \kappa\in(0,\infty) $, and $ \kappa=\infty $, respectively.
		\begin{enumerate}[(i)]
			\item $ \kappa=0 $: There exists a transition chemical potential  $ \mu^*\equiv\mu^*_0 \in[\mu_{\rm t},\mu_{\rm c}] $ such that
			$$
			\Mscr^{\ssup{0}}(\mu)=\begin{cases}  \{(\tilde{x}_1,0)\} &\mbox{ for } \mu<\mu^*,\\ \{(\tilde{x}_1,0),(\tilde{x}_2,\frac{\mu-a\tilde{x}_2}{a-b})\} & \mbox{ for } \mu=\mu^*,\\
			\{(\tilde{x}_2,\frac{\mu-a\tilde{x}_2}{a-b})\} & \mbox{ for } \mu>\mu^*\,.\end{cases}
			$$
			If $ \mu^*=\mu_{\rm c} $, there is always a unique zero, namely $ (\tilde{x}_1,0) $ for $ \mu \le \mu_{\rm c}$ with  $\tilde{x}_1(\mu^*)=\tilde{x}_2(\mu^*) $  and $ (\tilde{x}_2,\frac{\mu-\tilde{x}_2}{a-b}) $ for $ \mu>\mu_{\rm c} $.\\[1ex]
			
			\item $ \kappa\in (0,\infty) $: There exists a transition chemical potential $ \mu^*_\kappa $, namely $ \mu^*_\kappa=\mu^* $ for $ \mu_{\rm r}\le \mu^* $ with $ \mu^*\in[\mu_{\rm t},\mu_{\rm c}] $, see (i), and $ \mu^*_\kappa\in (\mu^*,\mu_{\rm r}) $ when $ \mu_{\rm r}>\mu^* $, 

			such that
			$$
			\Mscr^{\ssup{\kappa}}(\mu)=\begin{cases}  \{(\tilde{x}_1,0)\} &\mbox{ for } \mu<\mu^*_\kappa,\\ \{(\tilde{x}_1,0),(\tilde{x}_3,\kappa)\} & \mbox{ for } \mu=\mu^*_\kappa\\
			\{(\tilde{x}_3,\kappa)\} & \mbox{ for } \mu\in (\mu^*_\kappa,\mu_{\rm r}],\\ \{(\tilde{x}_2,\frac{\mu-a\tilde{x}_2}{a-b})\} & \mbox{ for } \mu \ge  \mu_{\rm r}\,.\end{cases}
			$$
			
			\medskip
			
			\item $ \kappa=\infty $: Then $ \Mscr^{\ssup{\infty}}(\mu)=\{(\tilde{x}_1,0)\} $ for all $ \mu\in\R $.
		\end{enumerate}
	\end{theorem}
	
	\medskip

	The zeroes in the previous theorem lead immediately to the following pressure representations We define the sub-critical and the super-critical pressure respectively as
	\begin{equation*}
	p_{\ssup{\rm sub}}(\beta,\mu)=\inf_{s<0}\Big\{\frac{(\mu-s)^2}{2a}+p_0(\beta,s)\Big\},\qquad
	p_{\ssup{\rm sup}}(\beta,\mu)=\sup_{s<0}\Big\{\frac{(\mu-a)^2}{2a}-\frac{s^2}{2b}+p_0(\beta,\mu)\Big\}.
	\end{equation*}
	
	\medskip

	\begin{cor}[\textbf{Pressure representation}]\label{Cor-1}
		For $ \kappa=0 $ we
		have
		\begin{equation*}
		\begin{aligned}
		p^{\ssup{0}}(\beta,\mu,\alpha)&=\begin{cases} p_{\ssup{\rm sub}}(\beta,\mu) & \mbox{ for } \mu\le \mu^*,\\ p_{\ssup{\rm sup}}(\beta,\mu) & \mbox{ for } \mu\ge \mu^*\,,\end{cases}
		\end{aligned}
		\end{equation*}
		and for $ \kappa\in (0,\infty) $,
		\begin{equation*}
		\begin{aligned}
		p^{\ssup{0}}(\beta,\mu,\alpha)&=\begin{cases} p_{\ssup{\rm sub}}(\beta,\mu) & \mbox{ for } \mu\le \mu^*_\kappa,\\ 
		p_{\ssup{\rm sub}}(\beta,\mu)+\mu k-\frac{(a-b)}{2} \kappa^2 & \mbox{ for } \mu\in [\mu^*_\kappa,\mu_{\rm r}],\\
		p_{\ssup{\rm sup}}(\beta,\mu) & \mbox{ for } \mu\ge \max\{\mu_{\rm r}, \mu^*\} \,,\end{cases}
		\end{aligned}
		\end{equation*}
		and for $ \kappa=\infty $, $ p^{\ssup{\infty}}(\beta,\alpha,\mu)=p_{\ssup{\rm sub}}(\beta,\mu) $ for all $ \mu\in\R $.
	\end{cor}

\medskip

	We have some control over the transition potentials $\mu^*_\kappa$, $ \kappa\in [0,\infty) $,  in  the following theorem. Denote
	$$
	\beta_{\rm t}(a,b,d) := \Big(\frac{a}{(4\pi)^{d/2}}\frac{b}{a-b}\zeta \big(\frac{d}{2}-1\big)\Big)^\frac{2}{d-2} \mbox { for } d\ge 5 \mbox{ and } C_d := \frac{(4\pi)^{d/2}}{a}\frac{a-b}{b} \mbox{ for  } d\ge 1.	
	$$

	\medskip
		
\begin{theorem}
		\label{thm:mutandmur}
		For $d=2$ and all $\beta>0$,
	$$
		\mu_{\rm t}\left(\beta\right) = \frac{a}{4\pi}\left(\left(1+C_2\right)\log \left(1+C_2\right) - C_2\log C_2\right)\frac{1}{\beta}.
$$
		Otherwise we describe the high and low temperature behaviours.
		
		\textbf{Low temperature:}
		\begin{align*}
		d\geq 5 &  \implies \mu_{\rm t}\left(\beta\right) = \mu_{\rm c}\left(\beta\right) \text{ for } \beta\geq\beta_{\rm t}, \mbox{ and, as  } \beta\to\infty, \\
		\mu_{\rm t}\left(\beta\right) & \sim \begin{cases} \mu_{\rm c}(\beta)\big(1 - \frac{1}{\zeta(2)}\ex^{(-C_4\beta)}\big) & \mbox{ for } d = 4,\\
		 \mu_{\rm c}(\beta)\Big(1 - \frac{\pi}{C_3\zeta(\frac{3}{2})} \frac{1}{\beta^{\frac{1}{2}}}\Big) & \mbox{ for } d=3,\\
		\frac{1}{2}\frac{a-b}{b}\frac{1}{\beta}\log \beta & \mbox{ for } d=1 \,. \end{cases} 
		\end{align*}

		\textbf{High temperature:}
		\begin{align*}
		d\geq 3 &\implies \mu_{\rm t} \sim \Big(1-\frac{d}{2}\Big)\frac{a-b}{b}\frac{1}{\beta}\log \beta  \text{ as } \beta\to 0,\\
		d=1 &\implies \mu_{\rm t} \sim \frac{a-b}{b}\Big(\frac{\sqrt{\pi}}{2C_1}\Big)^\frac{1}{3}\frac{1}{\beta^\frac{2}{3}} \text{ as } \beta\to 0.
		\end{align*}
		
		Furthermore, for $\kappa\in\left(0,+\infty\right)$, we have
	$$
		\mu_{\rm r}\left(\beta,\kappa\right) = \left(a-b\right)\kappa + ap'\left(\beta,-b\kappa\right).
	$$
	\end{theorem}

\bigskip	
	
	A second conclusion from our results above is the asymptotic limit for the pair empirical particle density.
	
	\begin{cor}[\textbf{Thermodynamic limit of the pair particle density}]\label{Cor-2}
		For every $ \delta > 0$, 
		$$
		\lim_{\L\uparrow\R^d}\,\Qr\left( \bM_\L\in B_\delta(x,y)\right) =1,
		$$
		for each unique minimiser $ (x,y) $ in Theorem~\ref{THM2}, that is, when either $ \mu\not=\mu^* $ in case $ \kappa=0 $, $ \mu\not=\mu^*_\kappa $ for $ \kappa\in (0,\infty ) $, and for all $ \mu\in\R $ when $ \kappa=\infty $. 
	\end{cor}

\medskip

	\begin{remark}
		\begin{enumerate}[(a)]

			\item  For the mean field model, i.e., for $ b\equiv 0 $, the paper \cite{BCMP05} derived similar results for a specific choice of the sequence $ m_\L $, namely $ m_\L\le L^2 $ when $ \abs{\L}=L^d $.  This corresponds to $\kappa=0 $ in our general setting if we have $d\geq 3$. \\[1ex]
			
			\item Our analysis above shows  that at the critical values of the chemical potential, $ \mu=\mu^* $ for $ \kappa=0 $ and $ \mu=\mu^*_\kappa $ when $ \kappa\in (0,\infty) $, we do not have a unique zero. Our large deviation analysis shows that for every $ \delta$-neighbourhood $ \Ucal_\delta $ of $ \Mscr^{\ssup{0}}(\mu^*) $ for $ \kappa=0 $  of $ \Mscr^{\ssup{\kappa}}(\mu^*_\kappa) $ when $ \kappa\in (0,\infty) $, respectively,
			$$
			\lim_{\L\uparrow\R^d}\,\Qr\left( \bM_\L\notin \Ucal_\delta^{\rm c}\right)=0. 
			$$
			Denoting $ (x^{\ssup{i}}_{\rm c},y^{\ssup{i}}_{\rm c}) $, $i=1,2 $, the zeroes at these critical points, the concentration of measure problem ask whether there are $ \lambda_i\in [0,1], i=1,2 $, such that $ \lambda_1+\lambda_2=1 $ and
			$$
			\lim_{\L\uparrow\R^d}\,\Qr\left( \bM_\L\in B_\delta(x^{\ssup{i}}_{\rm c},y^{\ssup{i}}_{\rm c})\right)=\lambda_i,\quad i=1,2.
			$$
			This requires finer asymptotic analysis going beyond the large deviation analysis studied here. We devote future work to analyse the fluctuation behaviour of our model, e.g.  similar to the analysis in \cite{CD14} for the ideal Bose gas, as well as  the concentration of measure problem at criticality. 
		\end{enumerate}
		\hfill$\diamond$
	\end{remark}

\bigskip

	A third observation from our results above is that the expected density of particles in unbounded cycles is related to the expected density of $ M_\L^{\ssup{2}} $ denoted
	\begin{equation*}
	\varrho^{\ssup{\kappa}}(\mu) :=\lim_{\L\uparrow\R^d} \Er\left[M_\L^{\ssup{2}}\right].
	\end{equation*}

	\begin{cor}[\textbf{Density in 'infinitely long cycles'}]\label{Cor-3}
		
		\begin{enumerate}[(i)]
			\item  $ \kappa = 0 $,
			$$
			\varrho^{\ssup{0}}(\mu) =\begin{cases} 0 &\mbox{ for }\mu<\mu^*,\\ \frac{\mu-a\tilde{x}_2(\mu)}{a-b} & \mbox{ for } \mu>\mu^*\,,\end{cases}
			$$ 
			and $ \varrho^{\ssup{0}}(\mu) $ is not continuous at $ \mu=\mu^* $ whenever  $ \mu^*<\mu_{\rm c} $, whereas $ \varrho^{\ssup{0}}(\mu) $ is continuous at $ \mu=\mu^* $ with $ \tilde{x}_1(\mu^*)=\tilde{x}_2(\mu^*) $ whenever  $ \mu^*=\mu_{\rm c} $         which is equivalent to $ p^{\ddprime}_0(\beta,0) \le \frac{a-b}{ab} $.\\[1ex]
			
			\item $\kappa\in (0,\infty) $,
			$$
			\varrho^\ssup{\kappa}(\mu)=\begin{cases} 0 & \mbox{ for } \mu<\mu^*_\kappa, \\ \kappa & \mbox{ for } \mu\in (\mu^*_\kappa,\mu_{\rm r}],\\
			\frac{\mu-a\tilde{x}_2(\mu)}{a-b} & \mbox{ for } \mu\ge \mu_{\rm r}\,, \end{cases}
			$$
			and $ \varrho^\ssup{\kappa}(\mu) $ is continuous at $ \mu=\mu_{\rm r} $ and discontinuous at $ \mu=\mu^*_\kappa $. \\[1ex]

			\item $ \kappa=\infty $, then $ \varrho^{\ssup{\infty}}(\mu) =0 $ for $ \mu\in\R $.
		\end{enumerate}
	\end{cor}
	
\medskip

	The reader maybe be tempted to identify the expect density $ \varrho^{\ssup{\kappa}}(\mu) $ as the density of particle in unbounded (`infinitely long cycles')  cycles which may corresponds to the density of the Bose-Einstein condensate. To answer this conjecture we first define the condensate density, a definition which goes back to \cite{Gir60}, and which is frequently used, e.g. in \cite{Lew86,BLP,BDLPb}.
	
\medskip

	For any $ K\in\N $ define $D_K $ as the random particle density of cycles with length greater than $K$,
	$$
	D_K(\omega):=\sum_{k>K} \, k \blambda_k\left(\omega\right)\,.
	$$
	Then the particle density in infinitely long cycles is defined as 
	\begin{equation*}
	\Delta^{\ssup{\kappa}}(\mu):=\lim_{K\to\infty} \lim_{\L\uparrow\R^d} \Er\left[ D_K\right].
	\end{equation*}

	\bigskip
	
	\begin{theorem}[\textbf{Condensate Density}]\label{THM3}
		\begin{enumerate}[(a)]
			\item Let $ \beta>0$, $\alpha<0$, $\mu\in \R $, and $ a>b>0 $. For $\kappa=0$ and $\kappa=\infty$,
			\begin{equation*}
			\Delta^{\ssup{0}}(\mu)=\varrho^{\ssup{0}}(\mu),\qquad
			\Delta^\ssup{\infty}\left(\mu\right) = \Big(\frac{\mu}{a}-\varrho_c\Big)_+.			\end{equation*}
			For $\kappa\in\left(0,\infty\right)$, there exists $\hat{\mu}^*_\kappa\in\left[\mu^*_\kappa,\mu_{\rm r}\right]$, such that
			\begin{equation*}
			    \Delta^\ssup{\kappa}\left(\mu\right) = \begin{cases}
			    \left(\frac{\mu}{a}-\varrho_c\right)_+ &\text{for }\mu < \hat{\mu}^*_\kappa\\
			   \varrho^\ssup{\kappa}\left(\mu\right)&\text{for }\mu > \hat{\mu}^*_\kappa.
			    \end{cases}
			\end{equation*}
			
			\item Let $ b\uparrow a $ with fixed $\beta>0$, $ \mu\in\R$, $a>0 $, and $ \kappa\in[0,\infty) $, then $\Delta^{\ssup{\kappa}}(\mu)\sim \big(\frac{\mu}{a-b}\big)_+$.\\[1ex]

			\item Let $ \mu\to\infty $, with fixed $\beta>0$, $ \mu\in\R$, $a>b>0 $, and $ \kappa\in[0,\infty) $, then $\Delta^{\ssup{\kappa}}(\mu)\sim \frac{\mu}{a-b}$.
			
		\end{enumerate}
		
	\end{theorem}

	\medskip

	\begin{remark}\label{rem-final}
	Since $m_\L > K$ eventually, it follows that $\Delta^\ssup{\kappa}\left(\mu\right) \geq \varrho^\ssup{\kappa}\left(\mu\right)$. However, if $\kappa>0$ we have parameter ranges in which this inequality is strict. This indicates that a positive condensate density is held on cycles whose lengths diverge slower than $m_\L$, and there therefore not directly affected by the counter-term. In the $\kappa=\infty$ case, $m_\L$ diverges too quickly to affect any positive density, see Remark~\ref{rem-THM1}, and we approach the mean field model. The case $ \kappa =\infty $ is special in the sense that the density of `infinitely long cycles' as the zero of the rate function vanishes but the above order parameter does not vanish when $ \mu\ge a\varrho_{\rm c} $.
	\hfill $ \diamond $
	\end{remark} 
	
	\bigskip

\section{Proofs}\label{sec-proofs}
The proof of the main large deviation theorem is in Section~\ref{Sec-LDP}, and all remaining proofs about the zeroes of the rate function and the condensate density are in Section~\ref{Sec-Variational}.	
	
\subsection{Proof of the large deviation principles, Theorem~\ref{THM1}}\label{Sec-LDP}
To prove the large deviation principle in Theorem~\ref{THM1},  we adapt and extend the ideas in \cite{BLP}. Our model is on spatial cycle structures and thus our method is  different as the so-called condensate resides in cycles of unbounded length. For the negative counter term in the Hamiltonian the long cycles are the  relevant ones, whereas in \cite{BLP} the energy indices with low values are relevant. Though technically slightly more challenging, our method allows us to investigate the different ways the counter term scales with the volume - in this way providing insight how the cycle condensate is scaled with the volume.	
The standard Varadhan Lemma approach for our model does not work directly as, due to the counter term, lower semi-continuity is missing. The general idea is to find lower and upper bounds for the Hamiltonian, and then prove large deviation principles for the two bounds individually. The final step is then to identify the two bounds. The proof for the upper bound is in  Section~\ref{Sec-upperLDP} and the one for the lower bound in Section~\ref{Sec-lowerLDP}.

\subsubsection{Large Deviation Upper Bound}\label{Sec-upperLDP}
We split the proof in two parts. In the first one we derive the large deviation principle for pair empirical particle density under the reference measure, i.e., with no interaction. In the second step we will apply Varadhan's lemma to a lower bound of the Hamiltonian.

\noindent \textbf{Step1:} Define the reference measure $ \nu_\L^{\ssup{2}}:=\Qr\circ\bM_\L^{-1} $. 
\begin{lemma}
For any $ \kappa\in[0,\infty] $ the limiting logarithmic moment generating function for $ (\nu_\L^{\ssup{2}})_{\L\Subset\R^d} $ is given by
$$
\bL(s,t)=\begin{cases} p_0(\beta,\alpha+s) & \mbox{for } s\le -\alpha \mbox{ and } t\le -\alpha,\\
+\infty &\mbox{for } s<-\alpha  \mbox{ or } t>-\alpha.
\end{cases}
$$
The Legendre-Fenchel transform is
$$
\bL^*(x,y)=\sup_{s\le 0}\left\{sx-p_0(\beta,s)\right\} -\alpha(x+y)+p_0(\beta,\alpha).$$
\end{lemma}
\begin{proofsect}{Proof}
\begin{equation*}
\bL(s,t) = \lim_{\L\uparrow\R^d}\frac{1}{\beta\abs{\L}}\log \nu_\L^{\ssup{2}}\left(\e^{\abs{\L}(sM_\L^{\ssup{1}}+tM_\L^{\ssup{2}})}\right)\\=\lim_{\L\uparrow\R^d}\Big(\sum_{k\leq m_\L}q_k\ex^{\beta(s+\alpha)k}+\sum_{k>m_\L} q_k\ex^{\beta( t+\alpha)k}\Big)-p_0(\beta,\alpha).
\end{equation*}
Given the properties of the weights $ q_k $, we immediately see convergence of the first term on the right hand side for any value of $ \kappa\in[0,\infty] $ towards $ p_0(\beta,\alpha+s) $ for $ (s,t)\in (-\infty,-\alpha]\times(-\infty,-\alpha] $. Furthermore, for any $ v>0 $,
$$
\lim_{\L\uparrow\R^d}\sum_{k\ge m_\L} q_k\ex^{vk}\ge \lim_{\L\uparrow\R^d} q_{m_\L}\ex^{v  m_\L}=+\infty.
$$
This gives divergence for $ s<-\alpha $ or $ t>-\alpha $. The Legendre -Fenchel transform is then
$$
\begin{aligned}
\bL^*(x,y)&= \sup_{(s,t)\in\R^2} \left\{sx+ty-\bL(s,t)\right\}=\sup_{s\le -\alpha}\left\{sx-p_0(\beta,s+\alpha)\right\}  +\sup_{t\le -\alpha}\left\{ty\right\}+p_0(\beta,\alpha)\\
&=\sup_{s\le 0}\left\{sx-p_0(\beta,s)\right\} -\alpha(x+y)+p_0(\beta,\alpha). 
\end{aligned}
$$
\qed
\end{proofsect}

Now we define
\begin{equation}
\label{idealUpper}
J_\kappa(x,y):=\begin{cases} \bL^*(x,y) &\mbox {for }(x,y)\in\K(\kappa),\\
+\infty &\mbox{otherwise.}\end{cases}
\end{equation}
and can easily obtain the large deviation upper bound.

\begin{lemma}
For any closed set $C\subset \R^2$,
\begin{equation*}
    \limsup_{\L\uparrow\R^d} \frac{1}{\abs{\L}}\log \nu_\L^{\ssup{2}}\left(C\right) \le -\inf_{(x,y)\in C}\; J_\kappa(x,y).
\end{equation*}
\end{lemma}
\begin{proofsect}{Proof}
The upper bound for closed sets $ C\subset \R^2 $,
$$
\limsup_{\L\uparrow\R^d} \frac{1}{\abs{\L}}\log \nu_\L^{\ssup{2}}\left(C\right) \le -\inf_{(x,y)\in C}\; \bL^*(x,y)
$$
follows for example with Baldi's Lemma (see \cite{DZ09}). The exponential tightness follows from the fact that the finite domain of $\bL^* $ contains a neighbourhood of the origin. We improve this bound for $\left(x,y\right)\not\in \K\left(\kappa\right)$ by noting that the image of $M^\ssup{2}_\L$ is $\left\{0,\frac{m_\L}{\abs{\L}},\frac{m_\L+1}{\abs{\L}},\frac{m_\L+2}{\abs{\L}},\ldots\right\}$. So if $E\subset\left(0,\kappa\right)$, then $\nu^\ssup{2}_\L\left(M^\ssup{2}_\L\in E\right)=0$ for all $\L$ sufficiently large.
\qed
\end{proofsect}

\noindent \textbf{Step 2:}

The crucial step is to bound Hamiltonian in \eqref{Def-H} from below. We have $\Hyp_\L(\omega)\ge \Hscr\circ\bM_\L^{\ssup{2}}(\omega)$, where 
\begin{equation}\label{D-Hscr}
\Hscr\colon\R^2\to\R,\quad (x,y)\mapsto \Hscr(x,y)=-(\mu-\alpha)(x+y)+\frac{a}{2}(x+y)^2-\frac{b}{2}y^2.
\end{equation}
\begin{lemma}\label{L-upper}
Let $ C\subset\R^2 $ be closed, then
$$
 \limsup_{\L\uparrow\R^d}\frac{1}{\beta\abs{\L}}\log\;\int_{\bM^{-1}_\L(C)}\,\ex^{-\beta \Hyp_\L(\omega)}\;\Qr(\d\omega)\le  -\inf_{(x,y)\in C}\,\left\{J_\kappa(x,y)+\Hscr(x,y)\right\},
$$
and 
$$
p^{\ssup{\kappa}}(\beta,\alpha,\mu)\le \sup_{(x,y)\in\K(\kappa)}\,\left\{ \mu(x+y)-\frac{a}{2}(x+y)^2+\frac{b}{2}y^2-f_0(\beta,x) \right\}.
$$
\end{lemma}
\begin{proofsect}{Proof}
$$
\int_{\bM^{-1}_\L(C)}\,\ex^{-\beta \Hyp_\L(\omega)}\;\Qr(\d\omega) \le \int_C\;\ex^{-\abs{\L}\Hscr(x,y)}\,\nu^{\ssup{2}}_\L(\dx,\dy).
$$
The function $ \Hscr $ in \eqref{D-Hscr} is continuous and bounded below, thus the statement is simply a matter of applying Varadhan's Lemma. Taking $ C=\R^2 $ gives the bound on the thermodynamic pressure by using the Legendre-Fenchel transform for the pressure.
\qed
\end{proofsect}
In Section~\ref{Sec-lowerLDP} we obtain the corresponding lower bound on the thermodynamic pressure, i.e., confirming \eqref{pressure(1)}. Then  Lemma~\ref{L-upper}, in conjunction with the pressure representation in \eqref{pressure(1)}, gives the large deviation upper bound
\begin{equation*}
\limsup_{\L\uparrow\R^d}\frac{1}{\beta\abs{\L}}\log \mu^{\ssup{2}}_\L(C)\le-\inf_{(x,y)\in C}\left\{I^\ssup{\kappa}(x,y)\right\},\qquad C\subset\R^2 \mbox{ closed}.
\end{equation*}

\subsubsection{Large Deviation Lower Bound}\label{Sec-lowerLDP}
The large deviation lower bound is more delicate as we shall find an upper bound on the energy, that is, the counter term cannot be replaced by the square of the sum of the single terms.  In the following steps we derive a more detailed splitting of the empirical particle density which is the novel step in this type of large deviation proofs. Our splitting is based on properties derived for the pair empirical particle density splitting  in Section~\ref{Sec-upperLDP}. The rate function $ J_\kappa $,  defined in \eqref{idealUpper}, is a good rate function which follows from the fact that the origin is in the interior of the domain where the limiting logarithmic moment generating function is finite as well as the fact that $ J_\kappa $ is convex and lower semi-continuous. We combine this with the local nature of the large deviation lower bound, that is, we will show that for any $ (x,y)\in\R^2 $,
\begin{equation}\label{LDP-lowerlocal}
\lim_{\delta\downarrow 0}\liminf_{\L\uparrow\R^d}\frac{1}{\beta\abs{\L}}\log\mu^{\ssup{2}}_\L(B_\delta((x,y))\ge - J_\kappa(x,y)-\Hscr(x,y)\,,
\end{equation}
where $ B_\delta((x,y)) $ is the open ball of radius $ \delta>0 $ around $ (x,y) $. In the first step we define the detailed splitting which is local as it depends on $ (x,y) $. In the second step we prove a large deviation lower bound  under the reference measure   for the new splitting of the empirical particle into a quadruple, and, using the derived upper bound on the energy, the lower bound for the quadruple splitting follows with Varadhan's Lemma. In a final step we employ the contraction principle to derive the lower bound in \eqref{LDP-lowerlocal}.

\medskip

\noindent \textbf{Step 1:  Splitting.}  Since $ J_\kappa +\Hscr $ has compact level sets, there exists $ (\tilde{x},\tilde{x})\in\K(\kappa) $ such that
$$
\inf_{(x,y)\in\R^2}\{J_\kappa(x,y)+\Hscr(x,y)\}=J_\kappa(\tilde{x},\tilde{y})+\Hscr(\tilde{x},\tilde{y}).
$$
Pick now $ (x,y)\in\K(\kappa) $, that is, $ y\in\{0\}\cup[\kappa,\infty) $ (for $ (x,y)\notin\K(\kappa) $ we obtain a trivial lower bound). The splitting depends on both $ (x,y) $ and $ (\tilde{x},\tilde{y}) $. Define $ r_\L:=\abs{\L}\tilde{y}\lor m_\L $, and then 
\begin{equation*}
\begin{aligned}
s_\L:=\begin{cases} (\abs{\L}y\lor m_\L)+1 & \mbox{ if } r_\L=\abs{\L}y \lor m_\L,\\
\abs{\L}y\lor m_\L & \mbox{ if } r_\L\not=\abs{\L}\lor m_\L.\end{cases}
\end{aligned}
\end{equation*}
Define the map 
$$
\bpi\colon\Omega\to\R^4_+,\quad \omega\mapsto\bpi(\omega)=(\bpi^{\ssup{1}}(\omega),\ldots,\bpi^{\ssup{4}}(\omega)),
$$
where
$$
\begin{aligned}
\bpi^{\ssup{1}}(\omega)&=\sum_{k=1}^{m_\L-1}\,k\blambda_k(\omega),\quad \bpi^{\ssup{2}}(\omega) =r_\L\blambda_{r_\L}(\omega),\quad
\bpi^{\ssup{3}}(\omega)=s_\L\blambda_{s_\L}(\omega),\\\bpi^{\ssup{4}}(\omega) &=\sum_{k\ge m_\L, k\not=r_\L,s_\L} k\blambda_k(\omega),
\end{aligned}
$$
the reference measure $ \nu_\L^{\ssup{4}}:=\Qr\circ\bpi^{-1} $, and the rate function 
$$
J^{\ssup{4}}_\kappa(x,y_1,y_2,z)=\begin{cases} f_0(\beta,x)-\alpha(x+y_1+y_2+z)+p_0(\beta,\alpha) & \mbox{for } (x,y_1,y_2,z)\in\bK,\\ +\infty & \mbox{otherwise,}\end{cases}
$$
where $ \bK:=\R\times\tilde{y}\N_0\times y\N_0\times(\{0\}\cup[\kappa,\infty)) $.

\noindent\textbf{Step 2: LDP lower bound for the quadruple splitting under the reference measure.}

\begin{lemma}
For any open set $ O\subset\R^4 $,
$$
\liminf_{\L\uparrow\R^d}\frac{1}{\beta\abs{\L}}\log \nu_\L^{\ssup{4}}(O)\ge -\inf_{(x,y_1,y_2,z)\in O} \,\left\{ J^{\ssup{4}}_\kappa (x,y_1,y_2,z)\right\}.
$$
\end{lemma}

\begin{proofsect}{Proof}
Pick $ (x,y_1,y_2,z)\in\bK $ (the other case gives trivial lower bound). The four particle densities $ \bpi^{\ssup{i}}, i=1,\dots,4 $, are independent and we derive the lower bound separately for each individual entry. Denote
$ \nu_{\L,i}^{\ssup{4}}, i=1,\ldots, 4 $, the marginals of $ \nu_\L^{\ssup{4}} $, and let $ \bL_i $ be  the limiting logarithmic moment generating functions for $ \nu_{\L,i}^{\ssup{4}} $ whose domains are strictly bounded by $ -\alpha $ and thus each contain a neighbourhood of the origin. The corresponding Legendre-Fenchel transforms are
\begin{equation*}
\bL^*_1(x)=f_0(\beta,x)-\alpha x +p_0(\beta,\alpha),\quad \bL^*_2(y_1)=-\alpha y_1,\quad \bL^*_3(y_2)=-\alpha y_2,\quad \bL^*_4(z)=-\alpha z.
\end{equation*}
We now derive for each marginal individual lower bounds.

\noindent \textbf{Marginal} $ \boldsymbol{\nu_{\L,1}^{\ssup{4}}} $: The function $ \bL^*_1 $ is strictly convex on $ [0,\varrho_{\rm c}) $. For $ x\in [0,\varrho_{\rm c}) $, we can proceed by standard G\"artner-Ellis type arguments by tilting the measure. Let $ \eta<0 $ be the unique solution to $ \eta x-p_0(\beta,\eta)=\sup_{s<0}\{sx-p_0(\beta,s)\} $, and define the tilted measure by
$$
\frac{\d \nu_{\L,\eta,1}^{\ssup{4}}}{\d\nu_{\L,1}^{\ssup{4}}}(x)=\exp\left(\beta\abs{\L}\left((\eta-\alpha)x-p_\L(\eta)+p_\L(\alpha)\right)\right),
$$
where $ p_\L(s):=\frac{1}{\beta}\sum_{k< m_\L} q_k\ex^{\beta sk} $. Taking the limits, we obtain
\begin{equation}\label{lower1}
\begin{aligned}
\lim_{\delta\downarrow 0}\liminf_{\L\uparrow \L}\frac{1}{\beta\abs{\L}}\log \nu_{\L,1}^{\ssup{4}}(B_\delta(x))\ge -f_0(\beta,x)+\alpha x-p_0(\beta,\alpha)+\lim_{\delta\downarrow 0}\liminf_{\L\uparrow\R^d}\frac{1}{\beta\abs{\L}}\log\nu_{\L,\eta,1}^{\ssup{4}}(B_\delta(x)).
\end{aligned}
\end{equation}
Now, using the strict convexity and the G\"artner-Ellis upper bound (see \cite[Theorem~2.3.6]{DZ09}), the last term on the right hand side of \eqref{lower1} vanishes as $ \nu_{\L,\eta,1}^{\ssup{4}}(B_\delta(x)) \to 1 $ as $ \L\uparrow\R^d $, and
\begin{equation}\label{lower2}
\lim_{\delta\downarrow 0}\liminf_{\L\uparrow \L}\frac{1}{\beta\abs{\L}}\log \nu_{\L,1}^{\ssup{4}}(B_\delta(x)) \ge -f_0(\beta,x)+\alpha x-p_0(\beta,\alpha)\,,\qquad\mbox{ for } x\in[0,\varrho_{\rm c}).
\end{equation}
For $ x>\varrho_{\rm c} $, the function $ \bL^*_1 $ is no longer strictly convex, and thus the standard argument fails to show that the last term on the right hand side of \eqref{lower1} vanishes. Our new method here is to directly use the Poisson nature of the distribution of the bosonic cycle counts to obtain individual estimates. Pick $ x>\varrho_{\rm c} $, and proceed as above to derive the lower bound in  \eqref{lower1}.  As we cannot exploit the strict convexity of $ \bL^*_1 $, we need to estimate the last term on the right hand side of \eqref{lower1} directly: pick $ r\in\N $, then eventually $ m_\L> r $. Denote $ \varrho_{\rm c}^\eta:=\sum_{k\in\N}kq_k\ex^{\beta \eta k} $,  $ \varrho_{{\rm c},r}^\eta:=\sum_{k< r} kq_k\ex^{\beta\eta k} $, and let $ \Qr_\eta $ be the distribution of the reference Poisson process with parameters $ \left\{q_k\ex^{\eta\beta k}\right\}_{k\in\N} $. Then for $ x>  \varrho_{{\rm c},r}^\eta $,
$$
\begin{aligned}
\nu_{\L,1}^{\ssup{4}}(B_\delta(x))\ge \Qr_\eta\Big(\sum_{k=1}^{r-1}k\blambda_k\in B_{\delta/2}(\varrho_{\rm c}^\eta)\Big)\Qr_\eta\Big(\abs{\L}\blambda_r=\floor{\abs{\L}/r(x-\varrho_{{\rm c},r}^\eta)}\Big)\Qr_\eta\Big(\sum_{k=r+1}^{m_\L-1}\blambda_k=0\Big).
\end{aligned}
$$
Now the mean and variance of $ \abs{\L}\sum_{k=1}^{r-1}k\blambda_k $ are equal to $ \abs{\L}\varrho_{{\rm c},r}^\eta $, so Chebyshev's inequality implies that
$$
\lim_{\L\uparrow\R^d} \Qr_\eta\Big( \sum_{k=1}^{r-1}k\blambda_k\in B_{\delta/2}(\varrho_{\rm c}^\eta)   \Big)\ge 1-\lim_{\L\uparrow\R^d}\frac{4\varrho_{{\rm c},r}^\eta}{\abs{\L}\delta^2}=1.
$$
Independence of the Poisson variables implies that
$$
\lim_{\L\uparrow\R^d}\frac{1}{\beta\abs{\L}}\log\Qr_\eta\Big(\sum_{k=r+1}^{m_\L -1} \blambda_k=0\Big)=-\lim_{\L\uparrow\R^d}\sum_{k=r+1}^{\m_\L-1}q_k\ex^{\beta \eta k} =-\sum_{k>r}q_k\ex^{\beta\eta k}.
$$
For the remaining factor we use Stirling's formula to get
$$
\begin{aligned}
\frac{1}{\beta\abs{\L}}\log\Qr_\eta\left( \abs{\L}  \blambda_r=\floor{\abs{\L}/r(x-\varrho_{{\rm c},r}^\eta)} \right) = & -\frac{(x-\varrho_{{\rm c},r}^\eta)}{r}\left(\log\frac{(x-\varrho_{{\rm c},r}^\eta)}{r} -\log q_r -1\right)\\ & +\eta(x-\varrho_{{\rm c},r}^\eta)-q_r\ex^{\beta \eta r}+o(1).
\end{aligned}
$$
Combing the individual estimates and taking $ r\to\infty $ gives 
$$
\liminf_{\L\uparrow\R^d}\frac{1}{\beta\abs{\L}}\log\nu_{\L,1}^{\ssup{4}}(B_\delta(x))\ge \eta(x-\varrho_{\rm c}^\eta)\ge \eta(x-\varrho_{\rm c}).
$$
Since $ \eta<0 $ can be chosen arbitrarily close to zero, we obtain \eqref{lower2} for all $x\ge 0 $. 

\noindent \textbf{Marginal} $ \boldsymbol{\nu_{\L,i}^{\ssup{4}}, i=2,3} $: If $ \tilde{y}=0 $, then
$$
\frac{1}{\beta\abs{\L}}\log \nu_{\L,2}^{\ssup{4}}(B_\delta(0))\ge \frac{1}{\abs{\L}}\log\Qr(\abs{\L}\blambda_{r_\L}=0)=-q_{r_\L}\ex^{\beta \alpha r_\L} =o(1).
$$
If $ \tilde{y}\ge \kappa $, then $ \frac{r_\L}{\abs{\L}}\to \tilde{y} $. Choose $ n\in\N_0 $ such that $ y_2=n\tilde{y} $. Then 
$$
\begin{aligned}
\frac{1}{\beta\abs{\L}}\log \nu_{\L,2}^{\ssup{4}}(B_\delta(n\tilde{y})) & \ge \frac{1}{\beta\abs{\L}}\log\Qr(\abs{\L}\blambda_{r_\L}=n)=\alpha n\frac{r_{\L}}{\beta\abs{\L}}+O(\frac{1}{\beta\abs{\L}}\log \abs{\L})+O(\frac{1}{\beta\abs{\L}}\log q_{r_\L})\\ & \quad +O(q_{r_\L}) =\alpha n \tilde{y} +o(1).
\end{aligned}
$$
Precisely the same argument applies for the case $ i=3 $.

\noindent \textbf{Marginal} $ \boldsymbol{\nu_{\L,4}^{\ssup{4}}} $:  If $ z=0 $,  let $ p_\L:=\min\{k\in\N\colon k\ge m_\L,k\not= r_\L,s_\L\} $. Then
$$
\frac{1}{\beta\abs{\L}}\log\nu_{\L,4}^{\ssup{4}}(B_\delta(0))\ge \frac{1}{\beta\abs{\L}}\log\Qr(\abs{\L}\blambda_{p_\L}=0)=o(1).
$$
If $ z\ge \kappa $, then let $ p_\L:=\min\{k\in\N\colon k\ge z\abs{\L}, k\ge  m_\L, k\not= s_\L,r_\L\} $. In particular, $ p_\L/\abs{\L}\to z $. Then
$$
\frac{1}{\beta\abs{\L}}\log\nu_{\L,4}^{\ssup{4}}(B_\delta(z))\ge \frac{1}{\beta\abs{\L}}\log\Qr(\abs{\L}\blambda_{p_\L}=1)=\alpha z +o(1).
$$

We finally  obtain for $ (x,y_1,y_2,z)\in\bK $,
$$
\lim_{\delta\downarrow 0}\liminf_{\L\uparrow\R^d}\frac{1}{\beta\abs{\L}}\log \nu_\L^{\ssup{4}}\left(B_\delta(x,y_1,y_2,z)\right)\ge -J^{\ssup{4}}_\kappa(x,y_1,y_2,z).
$$
by combining our  marginal estimates above.
\qed
\end{proofsect}

\noindent\textbf{Step 3: Upper bound for the energy and LDP lower bound for the quadruple splitting.}
Define 
$$ 
\Kscr(x,y_1,y_2,z)=-(\mu-\alpha)(x+y_1+y_2+z)+\frac{a}{2}(x+y_1+y_2+z)^2-\frac{b}{2}(y_1^2+y_2^2).
$$
Then,
\begin{equation}\label{upperboundenergy}
\Hyp_\L(\omega)\le \Kscr\circ \pi(\omega).
\end{equation}

\begin{lemma}\label{L-lower1}
Let $ O\subset\R^4 $ be open, then
$$
\liminf_{\L\uparrow\R^d}\frac{1}{\beta\abs{\L}}\log\,\int_{\bpi^{-1}(O)}\;\ex^{-\beta\Hyp_\L(\omega)}\,\Qr(\d\omega)\ge -\inf_{(x,y_1,y_2,z)\in O} \,\left\{ J^{\ssup{4}}_\kappa (x,y_1,y_2,z) +\Kscr(x,y_1,y_2,z)   \right\},
$$
and
$$
p^{\ssup{\kappa}}(\beta,\alpha,\mu)\ge p_0(\beta,\alpha)+\sup_{(x,y_1,y_2,z)\in \K(\kappa)}\left\{-J^{\ssup{4}}_\kappa(x,y_1,y_2,z)-\Kscr(x,y_1,y_2,z)\right\}.
$$
\end{lemma}

\begin{proofsect}{Proof}
Using \eqref{upperboundenergy},
$$
\int_{\bpi^{-1}(O)}\;\ex^{-\beta\Hyp_\L(\omega)}\,\Qr(\d\omega)\ge \int_O\;\ex^{-\beta\abs{\L}\Kscr(x,y_1,y_2,z)}\,\nu_\L^{\ssup{4}}(\d x,\d y_1,\d y_2,\d z).
$$
Then, noting that $ \Kscr $ is continuous, the statement is simply a matter of applying  Varadhan's Lemma. 
\qed
\end{proofsect}

\noindent\textbf{Step 4: Contraction.}

\begin{lemma}\label{L-lower2}
For any open set $ O\subset\R^2 $,
$$
\liminf_{\L\uparrow\R^d}\frac{1}{\beta\abs{\L}}\log\,\int_{\bM_\L^{-1}(O)}\;\ex^{-\beta \Hyp_\L(\omega)}\,\Qr(\d\omega) \ge -\inf_{(x,y)\in O}\left\{ J_\kappa(x,y)+\Hscr(x,y)\right\},
$$ and
$$
p^{\ssup{\kappa}}(\beta,\alpha,\mu)\ge p_0(\beta,\alpha)-\inf_{(x,y)\in\R^2}\left\{J_\kappa(x,y)+\Hscr(x,y)\right\}.
$$
\end{lemma}

\begin{proofsect}{Proof}
Define $ \hat{\bpi}\colon\R^4\to\R^2, (x,y_1,y_2,z)\mapsto (x,y_1+y_2+z) $, and note that $ \hat{\bpi} $ is continuous and $ \bM_\L=\hat{\bpi}\circ\bpi $. 
Then,
$$
\begin{aligned}
\inf_{x=X,y_1+y_2+z=Y}\left\{\left(J^{\ssup{4}}_\kappa+\Kscr\right)(x,y_1,y_2,z)\right\} &\ge f_0(\beta,X)-\mu(X+Y)+\frac{a}{2}(X+Y)^2-\frac{b}{2}Y^2+p_0(\beta,\alpha)\\
&=J_\kappa(X,Y)+\Hscr(X,Y),
\end{aligned}
$$
and, $\inf_{\R^4}\left\{J^{\ssup{4}}_\kappa+\Kscr\right\}\ge \inf_{\R^2}\left\{J_\kappa+\Hscr\right\}=(J_\kappa+\Hscr)(\tilde{x},\tilde{y}) $. Using 
$$ 
J_\kappa(\tilde{x},\tilde{y})+\Hscr(\tilde{x},\tilde{y})=J^{\ssup{4}}_\kappa(\tilde{x},\tilde{y},0,0)\ge \inf_{\R^4}\left\{J^{\ssup{4}}_\kappa+\Kscr\right\},
$$ we derive the lower bound for the pressure in Lemma~\ref{L-lower2}, and, with the corresponding upper bound in Lemma~\ref{L-upper}, we obtain  \eqref{pressure(1)} in Theorem~\ref{THM1}. To obtain the large deviation lower bound, denoted the rate function per Contraction principle $ \Jscr_\kappa(x,y)=\inf_{\hat{\bpi}^{-1}(x,y)}\{J^{\ssup{4}}_\kappa+\Kscr\}$. Then, for every open set $ O\subset\R^2 $, using the lower in Lemma~\ref{L-lower1},
$$
\liminf_{\L\uparrow\R^d}\frac{1}{\beta\abs{\L}}\log\,\int_{\bpi^{-1}\circ \hat{\bpi}^{-1}(O)}\;\ex^{-\beta\Hyp_\L(\omega)}\,\Qr(\d\omega)\ge -\inf_{\hat{\bpi}^{-1}(O)}\{J^{\ssup{4}}_\kappa+\Kscr\}=-\inf_{O}\{\Jscr_\kappa\},
$$
and, finally we derive \eqref{LDP-lowerlocal},
$$
\lim_{\delta\downarrow 0}\liminf_{\L\uparrow\R^d}\frac{1}{\beta\abs{\L}}\log \mu_\L^{\ssup{2}}(B_\delta(x,y))\ge -\Jscr_\kappa(x,y) \ge -(J^{\ssup{4}}_\kappa+\Kscr)(x,0,y,0)=-J_\kappa(x,y) -\Hscr(x,y),
$$
where we use the particular quadruple splitting in the last step.
\qed
\end{proofsect}

\subsection{Variational analysis}\label{Sec-Variational}
	
	\begin{proofsect}{Proof of Theorem~\ref{THM2}}
		In \cite{BLP} a related variational principle has been studied. We built on that and extend it to the whole  range of the parameter $ \kappa\in[0,\infty] $ to obtain an improvement on the results in \cite{BLP}. To obtain the zeroes of the rate function functions $ I^{\ssup{\kappa}} $ it suffices to minimise 
		\begin{equation*}
		F_\mu(x,y)=\begin{cases} f_0(\beta,x)-\mu(x+y)+\frac{a}{2}(x+y)^2 -\frac{b}{2} y^2 & \mbox{for }(x,y)\in\K(\kappa),\\
		+\infty & \mbox{otherwise.}\end{cases}
		\end{equation*}
		
		For $\kappa = +\infty$, we have restricted the $y$-component to $y=0$. Therefore we want to optimise $F\left(x,0\right)$. We search for stationary points, and find that they satisfy the equation $s_\beta\left(x\right) = \mu - ax$. Since $s_\beta$ is non-decreasing and $s_\beta\left(x\right)\to-\infty$ as $x\downarrow0$, and $\mu-a x$ is decreasing, there exists a unique stationary point - a minima - given by $\tilde{x}_1$.
		
		For $\kappa = 0$, we first fix $x\geq0$ and find an optimal choice of $y\in\R_+$, denoted $\tilde{y}\left(x\right)$. This gives $\tilde{y}\left(x\right) = \frac{1}{a-b}\left(\mu-ax\right)_+$, and the derivative of $F_\mu\left(x,\tilde{y}\left(x\right)\right)$ is zero precisely when
		\begin{equation*}
		s_\beta(x)=t_0(x):=\begin{cases} \mu-ax & \mbox{for } x\ge \frac{\mu}{a},\\ -\frac{b}{a-b}(\mu-ax) & \mbox{for } 0\le x\le \frac{\mu}{a}\,.\end{cases}
		\end{equation*}
		Now $x=\tilde{x}_1\left(\mu\right)$ is always a solution and gives a local minima for $\mu\leq \mu_c$, whilst $x=\tilde{x}_2\left(\mu\right)$ is a solution for $\mu\geq\mu_t$ and gives a local minima for $\mu>\mu_t$. These are the only local minima.
		
		Since $\tilde{x}_1\left(\mu\right) \geq \frac{\mu}{a}$ and $\tilde{x}_2\left(\mu\right) < \frac{\mu}{a}$, and $\tilde{x}_1\left(\mu\right)$ is the unique local minimiser for $\mu=\mu_t$ and $\tilde{x}_1\left(\mu\right)$ is the unique local minimiser for $\mu=\mu_c$,
        \begin{equation*}
            F_\mu\left(\tilde{x}_1\left(\mu_t\right),0\right) < F_\mu\left(\tilde{x}_2\left(\mu_t\right),\tilde{y}\left(\tilde{x}_2\left(\mu_t\right)\right)\right), \qquad F_\mu\left(\tilde{x}_1\left(\mu_c\right),0\right) > F_\mu\left(\tilde{x}_2\left(\mu_c\right),\tilde{y}\left(\tilde{x}_2\left(\mu_c\right)\right)\right),
        \end{equation*}
        and
        \begin{equation*}
            \frac{\d}{\d \mu}\left(F_\mu\left(\tilde{x}_1\left(\mu\right),0\right) - F_\mu\left(\tilde{x}_2\left(\mu\right),\tilde{y}\left(\tilde{x}_2\left(\mu\right)\right)\right)\right) = \frac{1}{a}\left(s_\beta\left(\tilde{x}_1\left(\mu\right)\right) - s_\beta\left(\tilde{x}_2\left(\mu\right)\right)\right) >0.
        \end{equation*}
        In taking the full $\mu$-derivative, we used the fact that $\tilde{x}_1\left(\mu\right)$ and $\tilde{x}_2\left(\mu\right)$ are stationary points. The inequality comes from $s_\beta$ being strictly increasing on $x\leq\varrho_c$. This tells us that there exists a $\mu^*\in\left[\mu_t,\mu_c\right]$ such that $\tilde{x}_1\left(\mu\right)$ is the minima for $\mu\leq \mu^*$ and $\tilde{x}_2\left(\mu\right)$ is the minima for $\mu\geq \mu^*$ - with simultaneous minima at $\mu=\mu^*$.
        
        Now suppose $\kappa\in\left(0,+\infty\right)$. If $\mu\geq\mu_r$ or $\mu\leq \mu^*$, then at least one global $\R^2_+$ minimiser exists in $\mathbb{K}$ and we are done. For $\mu\in\left(\mu^*,\mu_r\right)$, the quadratic form of $F$ in $y$ tells us that the $\mathbb{K}$ minimiser resides on $y=0$ or $y=\kappa$. Furthermore, by rearranging terms we find $F_\mu\left(x,\kappa\right) = F_{\mu-a\kappa}\left(x,0\right) - \kappa\mu + \frac{a-b}{2}\kappa^2$.
        Therefore $F_\mu\left(x,\kappa\right)$ is minimised at $x=\tilde{x}_1\left(\mu-a\kappa\right) = \tilde{x}_3\left(\mu\right)$. Now because $\left(\tilde{x}_1\left(\mu^*\right),0\right)$ is a global $\R^2_+$ minimiser at $\mu=\mu^*$ and $\left(\tilde{x}_3\left(\mu_r\right),\kappa\right) = \left(\tilde{x}_2\left(\mu_r\right),\kappa\right)$ is a global $\R^2_+$ minimiser at $\mu=\mu_r$,
        \begin{equation*}
            F\left(\tilde{x}_1\left(\mu^*\right),0\right) < F\left(\tilde{x}_3\left(\mu^*\right),\kappa\right), \qquad F\left(\tilde{x}_1\left(\mu_r\right),0\right) > F\left(\tilde{x}_3\left(\mu_r\right),\kappa\right).
        \end{equation*}
        Furthermore,
        \begin{equation*}
            \frac{\d}{\d \mu}\left(F\left(\tilde{x}_1\left(\mu\right),0\right) - F\left(\tilde{x}_3\left(\mu\right),\kappa\right)\right) = \frac{1}{a}\left(s_\beta\left(\tilde{x}_1\left(\mu\right)\right) - s_\beta\left(\tilde{x}_1\left(\mu-a\kappa\right)\right)\right) \geq 0,
        \end{equation*}
        since $s_\beta$ and $\tilde{x}_1$ are non-decreasing. Therefore there exists $\mu^*_\kappa\in\left(\mu^*,\mu_r\right)$ such that $\left(\tilde{x}_1\left(\mu\right),0\right)$ is a minimiser over $\mathbb{K}\left(\kappa\right)$ for $\mu\leq\mu^*_\kappa$ and $\left(\tilde{x}_3\left(\mu\right),\kappa\right)$ is a minimiser over $\mathbb{K}\left(\kappa\right)$ for $\mu\in\left[\mu^*_\kappa,\mu_r\right]$.
	\qed
	\end{proofsect}
	
	\begin{proofsect}{Proof of Corollary~\ref{Cor-1}}
		These representations of the pressure follow from substituting the minimisers from Theorem~\ref{THM2} into the pressure expression \eqref{pressure(1)} in Theorem~\ref{THM1}.
		\qed
	\end{proofsect}
	
	\begin{proofsect}{Proof of Theorem~\ref{thm:mutandmur}}
		We shall use the notation $g\left(u,s\right) = \sum^\infty_{k=1}k^{-s}\e^{u k}$, see \eqref{defBosefunctionapa}. First note that $\mu_{\rm t}=\mu_{\rm c}$ if and only if the infimum \eqref{Def-mu-t} is attained at $s=0$, which happens precisely when $ap_0''\left(\beta,0\right) \leq \frac{a-b}{b}$. This condition happens if and only if $d\geq 5$ and $\beta\geq\beta_{\rm t}$. In all other cases, there exists $\bar{\alpha}<0$ such that
		\begin{equation*}
		\mu_{\rm t} = a p_0'\left(\beta,\bar{\alpha}\right) - \frac{a-b}{b}\bar{\alpha} = \frac{a}{\left(4\pi\beta\right)^\frac{d}{2}}g\Big(\beta\bar{\alpha},\frac{d}{2}\Big) - \frac{a-b}{b}\bar{\alpha}.
		\end{equation*}
		Since $\bar{\alpha}$ is a stationary point, it is given as the unique solution to
		\begin{equation}
		\label{eqn:baralpha2}
		g\Big(\beta\bar{\alpha},\frac{d}{2}-1\Big) = \frac{\left(4\pi\right)^\frac{d}{2}}{a}\frac{a-b}{b}\beta^{\frac{d}{2}-1} = C_d\beta^{\frac{d}{2}-1}.
		\end{equation}
		
		To deal with $d=2$, note that from the infinite sum expression for $g\left(u,s\right)$ that for $u<0$ we have $g\left(u,0\right) = \frac{\e^u}{1-\e^u}$, and $g\left(u,1\right) = -\log\left(1-\e^u\right)$. We use these to solve for $\bar{\alpha}$ and then $\mu_{\rm t}$. For the remaining dimensions, we use \eqref{eqn:baralpha2} to determine whether $\beta\bar{\alpha}\to0$ or $\beta\bar{\alpha}\to-\infty$. Using $g\left(u,s\right)\sim\e^u$ as $u\to-\infty$ for all real $s$, and the $u\uparrow0$ asymptotic behaviour of $g\left(u,s\right)$ from Appendix~\ref{app-Bose}, we get the asymptotic behaviour of $\mu_{\rm t}\left(\beta\right)$.
		
		For $\mu_{\rm r}\left(\beta,\kappa\right)$, note that when restricted to $\left(0,\varrho_{\rm c}\right)$, the function $s_\beta:\left(0,\varrho_{\rm c}\right)\to\left(-\infty,0\right)$ is a strictly increasing continuous bijection. Furthermore we can re-write
$$
\mu_{\rm r} = \inf\left\{\sigma\geq\mu_{\rm t} : s_\beta\left(\frac{\sigma}{a}-\frac{\kappa}{a}(a-b)\right) \geq -\frac{b}{a-b}\left[\sigma - a\left(\frac{\sigma}{a}-\frac{\kappa}{a}(a-b)\right)\right] = -b\kappa\right\}
$$
		From the definition of $s_\beta(x)$ for $x\in(0,\varrho_{\rm c}]$, we have
$$
		s_\beta\left(\frac{\sigma}{a}-\frac{\kappa}{a}(a-b)\right) \geq  -b\kappa \iff \frac{\sigma}{a}-\frac{\kappa}{a}(a-b \geq p_0^\prime(\beta,-b\kappa)
		\iff \sigma \geq (a-b)\kappa + ap_0^\prime(\beta,-b\kappa).
$$
		\qed
	\end{proofsect}

	\begin{proofsect}{Proof of Corollary~\ref{Cor-2}}
		If $\left(x,y\right)$ is the unique minimiser of the large deviation rate function $I^\ssup{\kappa}$, then for $\delta>0$ we have $\inf\left\{I^\ssup{\kappa}\left(u,v\right) : \left(u,v\right) \not\in B^\delta\left(x,y\right)\right\} > 0$.
		Since the complement of the open ball $B^\delta\left(x,y\right)$ is closed, the large deviation principle tells us that $\Qr\left(\bM_\L \not\in B^\delta(x,y)\right)$ decays exponentially. Hence the required limit holds.
		\qed
	\end{proofsect}

	\begin{proofsect}{Proof of Corollary~\ref{Cor-3}}
		This follows from applying Corollary~\ref{Cor-2} with the unique minimisers found in Theorem~\ref{THM2}.
		\qed
	\end{proofsect}

	\begin{proofsect}{Proof of Theorem~\ref{THM3}}
	    Our strategy here is to use the large deviation techniques described previously to find $\lim_{\L\uparrow\R^d} \Er\left[ D_K\right]$, and then use analytic techniques to take the $K$-limit.
	    
		We begin by introducing the triple empirical particle density $\bM_{\L,K}\colon\Omega\to\R^3$,
		$\bM_{\L,K}(\omega):=\left(M_{\L,K}^{\ssup{1}},M_{\L,K}^{\ssup{2}},M_{\L,K}^{\ssup{3}}\right)$ where $M_{\L,K}^{\ssup{1}}=\sum_{k=1}^{K} k\blambda_k$, $M_{\L,K}^{\ssup{2}}=\sum^{m_\L}_{k=K+1}k\blambda_k$ and $M_{\L,K}^{\ssup{3}}=\sum_{k > m_\L}k\blambda_k$, distributed under the measure $\mu^\ssup{3}_{\L,K}:=\Pr_\L\circ \bM_{\L,K}^{-1}$. By repeating the arguments of Theorem~\ref{THM1}, we can show that $\left(\mu^\ssup{3}_{\L,K}\right)_{\L\Subset\R^d}$ satisfies the large deviation principle on $\R^3_+$ with rate $\beta\abs{\L}$ and good rate function
		\begin{equation*}
		I^{\ssup{\kappa}}_K(x,y,z)=\begin{cases} f^\ssup{1}_{K}(x) + f^\ssup{2}_{K}(y)-\mu(x+y+z)+\frac{a}{2}(x+y+z)^2-\frac{b}{2}z^2+p_{\kappa,K}(\beta,\alpha,\mu)&\mbox{if } z\not\in\left(0,\kappa\right),\\
		+\infty &\mbox{if } z\in\left(0,\kappa\right),\end{cases}
		\end{equation*}
		where $p_{\kappa,K}(\beta,\alpha,\mu)$ is the appropriate normalisation, $f^\ssup{1}_{K}\left(x\right) = \sup_{s\in\R}\left\{sx - \frac{1}{\beta}\sum_{j\leq K}q_j\e^{\beta sj}\right\}$ and $f^\ssup{2}_K\left(y\right) =\sup_{t\leq0}\left\{ty - \frac{1}{\beta}\sum_{j>K}q_j\e^{\beta tj}\right\}$. Note that $f^\ssup{1}_K\left(x\right)$ is convex, strictly increasing on $x\geq\varrho_c$ (and decreasing slower than linearly for $d=1,2$), and converges pointwise to $f_0\left(\beta,x\right)$. Also, $f^\ssup{2}_K\left(y\right)$ is convex, constant on $y\geq\sum_{k>K}kq_k$ and converges pointwise to $0$ (converges uniformly for $d\geq3$). From these properties, and since $\left(x,y,z\right)\mapsto-\mu(x+y+z)+\frac{a}{2}(x+y+z)^2-\frac{b}{2}z^2$ is a constant plus a positive definite quadratic form, we know that there are only finitely many global minimisers and that these are contained in some $K$-independent compact set. The pointwise convergence translates into uniform convergence on this compact set.
		
		Suppose $\kappa=\infty$. Then $I^\ssup{\kappa}_K$ is finite only if $z=0$, and therefore the question reduces to a mean field one. For $d=1,2$, the minima are eventually in a neighbourhood of the $\left\{y=z=0\right\}$ set, and so the condensate vanishes. For $d\geq 3$, the pointwise limit has a single global minimiser, $\left(\tilde{x}_1,0,0\right)$, for $\mu\leq\mu_c$ and uncountably many for $\mu>\mu_c$. These minimisers are the convex combinations of $\left(\frac{\mu}{a},0,0\right)$ and $\left(\varrho_c,\frac{\mu}{a}-\varrho_c,0\right)$. However, because we know that $f^\ssup{1}_K\left(x\right)$ is strictly increasing on $x\geq \varrho_c$ for all $K$, we know that the minimisers are eventually in any neighbourhood of $\left(\varrho_c,\frac{\mu}{a}-\varrho_c,0\right)$.
		
		For finite $\kappa$, we contract the $y$ and $z$ components to compare with $I^\ssup{\kappa}\left(x,y\right)$. After contracting $\left(x,y,z\right)\mapsto\left(x,y+z\right)$, we get the rate function
		\begin{equation*}
		    J^\ssup{\kappa}_K\left(x,y\right)= f^\ssup{1}_K\left(x\right) - \mu\left(x+y\right) + \frac{a}{2}\left(x+y\right)^2 + \inf_{Y+Z=y,Z\not\in\left(0,\kappa\right)}\Big\{f^\ssup{2}_K\left(Y\right) - \frac{b}{2}Z^2\Big\}.
		\end{equation*}
		This has the pointwise limit
		\begin{equation*}
		    J^\ssup{\kappa}_\infty\left(x,y\right)= f_0\left(\beta,x\right) - \mu\left(x+y\right) + \frac{a}{2}\left(x+y\right)^2 -\frac{b}{2}y^2\mathds{1}_{\left\{y\geq\kappa\right\}}.
		\end{equation*}
		Note $I^\ssup{\kappa}\left(x,y\right) \geq J^\ssup{\kappa}_\infty\left(x,y\right) \geq I^\ssup{0}\left(x,y\right)$, with equality precisely on $\mathbb{K}\left(\kappa\right)$. Therefore if $I^\ssup{0}$ has a global minimiser in $\mathbb{K}\left(\kappa\right)$, then this is a global minimiser of $J^\ssup{\kappa}_\infty$. This proves the $\kappa = 0$ case, and the $\kappa\in\left(0,\infty\right)$ case for $\mu\leq\mu^*$ or $\mu\geq\mu_{\rm r}$.
		
		By fixing $x$ and taking the $y$-partial derivatives of $J^\ssup{\kappa}_\infty$, we find that the optimal choice of $y$ is $\frac{\mu}{a}-x$ or $\kappa$ if $\frac{\mu-ax}{a-b}<\kappa$, and $\frac{\mu-ax}{a-b}$ otherwise. If $\mu\geq\mu_{\rm r}$, then $y=\frac{\mu-ax}{a-b}$ produces the global minimiser. For $\mu\in\left(\mu^*,\mu_{\rm r}\right)$ we need to compare minimisers along $y=\frac{\mu}{a}-x$ and $y=\kappa$. This is a similar process to the comparison of $y=0$ and $y=\kappa$ we performed in the proof of Theorem~\ref{THM2}. Furthermore, because $J^\ssup{\kappa}_\infty\left(x,\frac{\mu}{a}-x\right) < J^\ssup{\kappa}_\infty\left(x,0\right)$, we know that the transition occurs at a higher value of $\mu$.
		
		The asymptotics follow from inspecting the behaviour of $\varrho^\ssup{\kappa}\left(\mu\right)$ in the respective limits. In particular note that $\mu_{\rm r}< \mu_{\rm c}$ eventually as $b\uparrow a$.
		\qed
	\end{proofsect}

	\section*{Appendices}
	\begin{appendices}
		\section{Reference process  for Dirichlet  boundary conditions}\label{app-A}
		
		\noindent For Dirichlet boundary condition, one restricts the Brownian bridges to not leaving the set $\L$, that is,  one replaces $ \Ccal_{k,\L} $ by  the space $  \Ccal^{\ssup{\rm Dir}}_{k,\L} $ of continuous functions in $ \L $ with time horizon $ [0,k\beta] $.  Consider the measure
	$$
		\mu^{\ssup{{\rm Dir},k\beta}}_{x,y}(A)=\frac{\P_x(B\in A;B_{k\beta}\in\d y)}{\d y},\qquad A\subset\Ccal^{\ssup{\rm Dir}}_{k,\L}\mbox{ measurable},
$$
which has total mass $ g_{k\beta}^{\ssup{\rm Dir}}(x,y)=\mu_{x,y}^{\ssup{\rm Dir,k\beta}}(\Ccal^{\ssup{\rm Dir}}_{k,\L}) $.
For  Dirichlet boundary conditions \eqref{q*def} is replaced by
\begin{equation*}
\overline{q}^{\ssup{\rm Dir}}=\sum_{k\in\N} q_k^{\ssup{\rm Dir}},\qquad\mbox{ with } q_k^{\ssup{\rm Dir}}=\frac{1}{k|\L|}\int_\L\d x\,g_{k\beta}^{\ssup{\rm Dir}}(x,x).
\end{equation*}
		Note that these weight depends on $ \L $, see \cite{D2019}. We introduce the Poisson point process $\omega_{\rm P} = \sum_{x \in \xi_{ \rm P}} \delta_{(x,B_x)}$ on $\L\times E^{\ssup{\rm Dir }}$  with intensity measure $ \nu^{\ssup{\rm Dir}} $ whose projections on $ \L\times \Ccal_{k,\L}^{\ssup{\rm Dir}} $ with $k\leq \floor{\L}$ are equal to $ \nu_k^{\ssup{\rm Dir}}(\d x,\d f)=\frac{1}{k}\Leb_\L(\d x)\otimes\mu_{x,x}^{\ssup{{\rm Dir},k\beta}}(\d f)$ and are zero on this set for $k>N$. We do not label $\omega_{\rm P}$ nor $\xi_{ \rm P}$ with the boundary condition nor with $N$; $\xi_{ \rm P}$ is a Poisson process on $\L$ with intensity measure $\overline q^{\ssup{\rm Dir}}$ times the restriction $\Leb_\L$ of the Lebesgue measure to $\L$. By $\Qr^{\ssup{\rm Dir}}$ and $ \Er^{\ssup{\rm Dir}}$ we denote probability and expectation with respect to this process. Conditionally on $\xi_{ \rm P}$, the lengths of the cycles $B_x$ with $x\in\xi_{ \rm P}$ are independent and have distribution $(q_k^{\ssup{\rm Dir}}/\overline q^{\ssup{\rm Dir}})_{k\in\{1,\dots,\floor{\L}\}}$; this process has only marks with lengths $\leq \floor{\L}$. A cycle $B_x$ of length $k$ is distributed according to
		$$
		\P_{x,x}^{\ssup{{\rm bc},k\beta}}(\d f)=\frac{\mu_{x,x}^{\ssup{{\rm bc},k\beta}}(\d f)}{g^{\ssup{\rm bc}}_{k\beta}(x,x)}.
	$$

		The above representations allows us to prove the our large deviation principles as well as the variational analysis  for Dirchlet  boundary conditions. For details we refer to \cite{ACK} where these arguments are presented in detail. The independence of the thermodynamic limit of  pressure in Theorem~\ref{THM1} follows using either the arguments in \cite{ACK} or in \cite{R71,BR97}.
		
		\section{Bose function}\label{app-Bose}
		The Bose functions are poly-logarithmic functions defined by 
		\begin{equation}\label{defBosefunctionapa}
		g(n,\alpha):=\Li_n(\ex^{-\alpha})=\frac{1}{\Gamma(n)}\int_0^\infty\frac{t^{n-1}}{{\rm e}^{t+\alpha}-1}\,\d t=\sum_{k=1}^\infty k^{-n}{\rm e}^{-\alpha k}\quad\mbox{ for all } n \mbox{ and } \alpha>0,
		\end{equation}
		and also for $ \alpha=0 $ and $ n>1 $. In the latter case,
		\begin{equation}\label{zeta}
		g(n,0)=\sum_{k=1}^\infty k^{-n}=\zeta(n),
		\end{equation}
		which is Riemann zeta function. The behaviour of the Bose functions about $ \alpha=0 $ is given by
		\begin{equation}\label{boseexp}
		g(n,\alpha)=\begin{cases}
		\Gamma(1-n)\alpha^{n-1}+\sum_{k=0}^\infty\zeta(n-k)\frac{(-\alpha)^k}{k!} &\text{for } n\not= 1,2,3,\ldots, \\[1.5ex]
		\frac{(-\alpha)^{n-1}}{(n-1)!}\left[-\log\alpha +\sum_{m=1}^{n-1}\frac{1}{m}\right]+\sum_{\heap{k=0}{k\not= n-1}}\zeta(n-k)\frac{(-\alpha)^k}{k!} &\text{for } n\in\N.                   
		\end{cases}
		\end{equation}
		At $ \alpha=0 $, $ g(n,\alpha) $ diverges for $ n\le 1 $, indeed for all $n$ there is some kind of singularity at $ \alpha=0 $, such as a branch point. For further details see \cite{Gram25}. The expansions \eqref{boseexp} are in terms of $ \zeta(n) $, which for $ n\le 1 $ must be found by analytically continuing \eqref{zeta}. With the asymptotic properties of the zeta function it can be shown that the $k$ series in \eqref{boseexp} are convergent for $ |\alpha|<2\pi $. Consequently \eqref{boseexp} also represents an analytic continuation of $ g(n,\alpha) $ for $ \alpha<0 $. When $ \alpha\gg 1 $ the series \eqref{defBosefunctionapa} itself is rapidly convergent, and as $ \alpha\to\infty $, $ g(n,\alpha)\sim\ex^{-\alpha} $ for all $n$. 
%
%

	\end{appendices}

\end{document}